# Origin of Europa and the Galilean Satellites

**Robin M. Canup and William R. Ward**

*Southwest Research Institute*

Europa is believed to have formed near the very end of Jupiter's own accretion, within a circumplanetary disk of gas and solid particles. We review the formation of the Galilean satellites in the context of current constraints and understanding of giant planet formation, focusing on recent models of satellite growth within a circumjovian accretion disk produced during the final stages of gas inflow to Jupiter. In such a disk, the Galilean satellites would have accreted slowly, in more than $10^5$ yr, and in a low pressure, low gas density environment. Gravitational interactions between the satellites and the gas disk lead to inward orbital migration and loss of satellites to Jupiter. Such effects tend to select for a maximum satellite mass and a common total satellite system mass compared to the planet's mass. One implication is that multiple satellite systems may have formed and been lost during the final stages of Jupiter's growth, with the Galilean satellites being the last generation that survived as gas inflow to Jupiter ended. We conclude by discussing open issues and implications for Europa's conditions of formation.

# 1.    INTRODUCTION

Europa belongs to the family of four Galilean satellites, which each have similar masses, and orbits that are prograde, nearly circular, and approximately co-planar. These basic properties suggest that the Galilean satellites formed contemporaneously within a shared circumplanetary disk orbiting within Jupiter's equatorial plane. The conditions in this early disk established the satellites' initial masses, compositions, and thermal and orbital states. Observed properties of the Galilean satellites can provide constraints on this formation environment, which in turn can provide clues to the conditions of Jupiter's formation.

Regular satellites originate from material in circumplanetary orbit, which can arise from a number of diverse processes (*e.g.*, *Pollack et al.*, 1991). A large collision into a solid planet can produce an impact-generated disk or in some cases an intact satellite (*e.g.*, *Canup*, 2004, 2005). Although an impact origin is typically implicated for the Earth's Moon and Pluto's Charon, this seems an unlikely explanation for the Galilean satellites given Jupiter's small ~ 3° obliquity. Collisions between objects in solar orbit that occur relatively close to a planet can also leave material in planetary orbit, in a process known as co-accretion (*e.g.*, *Safronov et al.*, 1986). Forming the Galilean satellites by co-accretion is quite difficult, primarily because the rate of such collisions is generally low and captured material tends to not have enough net angular momentum to account for the current satellite orbits (*Estrada and Mosqueira*, 2006).

For gas planets, two processes are thought to be effective means for producing circumplanetary disks of gas (H + He) and solids (rock + ice). When a gas giant undergoes runaway gas accretion, it expands to nearly fill the entire region within which its gravity dominates over that of the Sun, known as the planet's Hill sphere. For Jupiter's distance and current mean density, the Hill sphere of a planet is more than 700 times the planet's radius. A forming gas planet will begin to contract within its Hill sphere once the rate of its gas accretion can no longer compensate for the increasing rate of its gravitational contraction due to the planet's growing mass and luminosity. Accreted gas delivers angular momentum to the planet, and as the spinning planet contracts, conservation of angular momentum dictates that its rate of rotation increases. When this rate reaches the critical rate for rotational instability, the planet's outer equatorial layers begin to shed, forming a so-called "spin-out" disk. A second means for forming a disk occurs if the solar nebula is still present during the contraction phase, so that while the planet contracts, it is also continuing to accrete gas. As the planet becomes smaller and smaller, at some point the gas inflowing from solar orbit contains too much angular momentum to allow it to fall directly onto the planet. Instead, gas flows into orbit around the planet and directly creates a disk (*e.g.*, Figure 1). Such a structure is known in astrophysical contexts as an "accretion" disk.

Whether the Galilean satellites formed in a spin-out or an accretion disk depends on the relative timing of Jupiter's contraction and the dispersal of the solar nebula. Recent models (*e.g.*, *Papaloizou and Nelson*, 2005) predict that a Jupiter-mass planet will contract to within the current orbit of Io in less than a million years, while estimates of the lifetime of the solar nebula are longer, $\sim 10^6$ to $10^7$ yr (*e.g.*, *Haisch et al.*, 2001). Thus it appears probable that an accretion disk would have predominated during the final stages of Jupiter's formation.

The implication is that the Galilean satellites formed in a primordial gas disk surrounding Jupiter, some millions to perhaps ten million years after the origin of the first solar system solids. The Jovian circumplanetary disk has often been viewed as analogous to the circumsolar disk, and in many respects the growth of satellites within this "mini solar system" would proceed similarly to the well-studied growth of solid planets from the nebular disk. Collisions between orbiting objects would lead to growth of progressively larger bodies. Interactions between growing satellites and the gas disk would modify their orbits, generally causing them to become more circular and co-planar, and then on longer timescales to spiral inward toward the planet. However, a fundamental difference between satellite and planet growth is that the satellite formation environment around a gas planet is coupled to and regulated by the evolution of the planet and the surrounding solar nebula. This leads to disk conditions that are not adequately described as simply a scaled-down version of the solar nebula, a theme that will be emphasized here.

In this chapter, we focus on a relatively recent model (*Canup and Ward*, 2002, 2006; *Ward and Canup*, in preparation) in which the Galilean satellites form within a gaseous accretion disk produced at the very end of Jupiter's own accretion (*e.g.*, Figure 2). In this construct, the disk is actively supplied by an ongoing inflow of gas and small solids to circumjovian orbit as satellites are forming. Disk conditions evolve in a quasi-steady state fashion in response to the evolution of Jupiter's own state and the rate of its gas accretion. This leads to lower disk gas densities and/or temperatures at the time of Galilean satellite formation than assumed in prior works. Within the disk, orbiting solids accumulate into satellites. Satellite growth rates are governed by the rate of delivery of solid particles from solar orbit, which leads to orders-of-magnitude slower satellite accretion than typically predicted by models that consider a static rather than an actively supplied disk. We envision that Jupiter had a circumplanetary disk for an extended time period, and that it is probable that a total mass of nebular gas and solids much larger than the amount needed to form the Galilean satellites would have been processed through this disk. As satellites grow to masses comparable to those of the Galilean satellites, their gravitational interactions with the gas disk can cause their orbits to spiral inward towards the planet. Earlier generations of satellites at Jupiter may have formed and been lost to collision with Jupiter, with the Galilean satellites we see today representing the last surviving generation of Jovian satellites that formed as gas inflow to the planet ended.

In Section 2, we begin by outlining key constraints on Galilean satellite origin that arise from the properties of the satellites themselves. We then describe the development and expected characteristics of a circumplanetary disk in the context of Jupiter's mass accretion and contraction (Section 3). In Section 4 we discuss the expected conditions within the disk at the time of the Galilean satellites' formation, as well as models of the growth (and loss) of large satellites within such a disk. Section 5 reviews primary outstanding issues and areas requiring future research, and in Section 6 we outline potential implications for Europa's origin.

The formation of the Galilean satellites has been discussed in many other review chapters, including *Stevenson et al.* (1986), *Coradini et al.* (1989), *Pollack et al.* (1991), *Peale* (1999), *Lunine et al.* (2004), *McKinnon* (2007), *Peale* (2008) and the *Estrada et al.* chapter in this volume. The reader is referred to these reviews for other viewpoints on this topic.

## 2. CONSTRAINTS

A successful origin model must be consistent with basic properties of the Galilean satellites (see Table 1).

### 2.1. Satellite masses

The Galilean satellites have similar masses, increasing by only about a factor of three between Europa and Ganymede. Together, the four satellites comprise a total satellite system mass, $M_T = 3.9 \times 10^{26}$ g, which is a small fraction of Jupiter's mass, $M_J = 1.9 \times 10^{30}$ g, with $(M_T/M_J) = 2.1 \times 10^{-4}$. This fraction is nearly identical to the analogous quantity for the Saturnian satellite system ($2.5 \times 10^{-4}$). Current interior models estimate that Jupiter and Saturn each contain substantial quantities of heavy elements other than H and He, with Jupiter having ~ 10 to 40 Earth masses ($M_\oplus$) in heavy elements and Saturn having ~ 2 to $8M_\oplus$ (*e.g.*, *Guillot*, 2005). Yet their largest satellites are all relatively small, containing within a factor of two of the mass of the Moon, which contains a much larger fraction of the Earth's mass ($\approx 0.01M_\oplus$). This suggests that processes that formed the gas planet satellites limited—and possibly even selected for—a maximum size to which the satellites could grow.

Assuming that the Galilean satellites formed from solar composition material (*i.e.*, that contains a gas-to-solids ratio of about $10^2$), a minimum mass of ~ $0.02M_J$ in gas and solids is required to produce them in the limit of perfectly efficient accretion. Historically, it is assumed that the satellites formed in a disk containing this amount of material spread over an area extending from approximately the planet's surface to the orbit of the outermost satellite (*e.g.*, *Lunine and Stevenson*, 1982; *Coradini and Magni*, 1984; *Coradini et al.*, 1989; *Pollack et al.*, 1991). Such a disk is referred to as the "minimum mass sub-nebula", or MMSN, in direct analogy to the "minimum mass nebula" defined for the circumsolar disk (*e.g.*, *Hayashi et al.*, 1985). The corresponding MMSN gas surface density is of order $\sigma_{g,MMSN} \sim 0.02 M_J / [\pi(30R_J)^2] \sim$ few $\times 10^5$ g/cm$^2$, where $R_J = 7.15 \times 10^4$ km is Jupiter's radius. A variant on the MMSN has recently been advocated by *Mosqueira and Estrada* (2003; see also chapter by *Estrada et al.*), who invoke a similarly high gas density disk interior to ~ $15R_J$, accompanied by an outer, lower density disk (with $\sigma_g \sim 10^3$ to $10^4$ g/cm$^2$) in the region exterior to Ganymede's orbit. In general, the MMSN construct implies a gas-rich environment for Galilean satellite formation, in which circumplanetary disk gas surface densities are many orders-of-magnitude higher than those in the surrounding solar nebula.

Considering a pre-existing disk as an initial condition reflects an implicit assumption that the disk forms quickly relative to the time it takes the objects within it to grow. For the circumsolar disk this assumption is probably valid: the timescale over which the disk forms due to molecular cloud core collapse (~ $10^4$ to $10^6$ yr, *e.g., Myers et al.*, 2000) is likely shorter than the timescale for planets to accrete (~ $10^6$ to $10^8$ yr, *e.g. Nagasawa et al.*, 2007; *Lissauer and Stevenson*, 2007). However for a circumplanetary disk, the timescale over which the disk forms may be comparable to the formation time of the giant planet itself (~ $10^5$ to $10^7$ yr, *e.g.*, *Lissauer and Stevenson*, 2007), and this is much longer than the timescale for satellite growth predicted within an MMSN-type disk ($\leq 10^3$ yr, *e.g.*, *Canup and Ward, 2002*). In such a situation, invoking a pre-existing MMSN disk is not appropriate, because the disk's formation would be ongoing as the satellites grow. While it is then still the case that a minimum mass of $0.02M_J$ in solar composition material must be delivered to the disk to produce the Galilean satellites, it need not be present in the disk all at one time (*Canup and Ward*, 2002). In a dynamically evolving disk, the instantaneous disk mass and surface density could be orders-of-magnitude less than that of the MMSN, so long as an appropriate total mass is processed through the disk during its lifetime.

To summarize, a minimum total mass in solar composition material ~ $0.02M_J$ (corresponding to a minimum solid mass of ~ $2 \times 10^{-4}M_J$) must be delivered to circumjovian orbit to produce the Galilean satellites. The corresponding protosatellite disk surface density then depends on whether the disk is considered to be a closed system with a fixed total mass (as assumed in the MMSN models) or one that is actively supplied as the satellites grow (as we advocate here).

## 2.2 Compact radial scale

The Galilean satellites are located between about 6 and 26 Jovian radii from the center of the planet. This is a quite compact configuration compared to the maximum orbital radius for gravitationally bound circumplanetary material, which is comparable to Jupiter's Hill radius, $R_{H,J} \equiv a_J (M_J / 3M_\odot)^{1/3} \approx 744 R_J$, where $a_J$ = 5.2 AU = $7.78 \times 10^{13}$ cm is Jupiter's semi-major axis and $M_\odot = 1.99 \times 10^{33}$ g is the Sun's mass. While Jupiter's irregular satellites have much more extended orbits, their total mass is negligible compared to that of the four large satellites and their capture may have occurred much later (*e.g.*, *Jewitt and Haghighipour*, 2007). Primordial process(es) that supplied the bulk of protosatellite material only resulted in large satellites on orbits relatively close to Jupiter, and this trend is also seen in the regular satellites of Saturn and Uranus.

## 2.3 Compositional gradient

The decrease in mean satellite density with increasing distance from Jupiter (Table 1) is associated with a progressive increase in satellite ice mass fraction. Innermost Io is presently anhydrous, and its 3.53 g/cm$^3$ density suggests an interior composed of rock and metal. Europa, with a density of 2.99 g/cm$^3$, is thought to contain about 8% ice and water by mass, with the remainder rock (*e.g.*, *Schubert et al.*, 2004). Outer Ganymede and Callisto have similar densities (1.94 and 1.83 g/cm$^3$, respectively), implying mixed ice/rock compositions and rock mass fractions ~ 0.55 for Ganymede and ~ 0.44 for Callisto (*McKinnon*, 1997). The compositional gradient seen in the Galilean satellites has traditionally been viewed as evidence that the satellites formed in a disk whose temperature decreased with orbital distance from the planet, with the ice-rich compositions of Ganymede and Callisto implying that the stability line for ice during the bulk of Galilean satellite growth was near the distance at which Ganymede formed (*e.g.*, *Lunine and Stevenson*, 1982). This distance could be somewhat exterior to Ganymede's current orbital radius due to inward satellite migration (see § 4.3.4). Europa's small ice content would then imply that the ice stability line moved closer to Jupiter during the final stages of satellite growth, allowing limited ice accretion onto this satellite.

The above interpretation assumes that the current satellite densities reflect their initial compositions. However it is possible—at least based on purely energetic arguments—that Europa could have begun with a more ice-rich composition, due to the potential for volatile loss associated with tidal dissipation over its history in the Laplace resonance. The current total heat flow from Europa has not been measured, but models predict $F_t \sim 6 \times 10^{18}$ erg/s to $3 \times 10^{19}$ erg/s (*e.g.*, *Hussmann et al.*, 2002; *Moore*, 2006). If heat generated by tidal dissipation were balanced by radiation from a uniform temperature surface, the implied surface temperature is only $T \sim \left[ F_t / (4\pi R_S^2 \sigma_{SB}) \right]^{1/4} \sim 30$ K $\left( F_t / 10^{19} \text{ erg/sec} \right)^{1/4}$, where $R_S = 1561$ km is Europa's radius and $\sigma_{SB}$ is the Stephan-Boltzman constant, implying minimal sublimation loss of ice. Alternatively, warm tidally-melted material could periodically erupt to the satellite's surface, with most of the energy lost through localized "hot-spots". In the limits that all tidally generated heat went into vaporizing water and that Europa has occupied the Laplace resonance for $4.5 \times 10^9$ yr (as predicted by some models, including *Peale and Lee*, 2002), a mass in ice equal to Europa's current mass ($M_S = 4.8 \times 10^{25}$ g) could have been lost over its history, because $(10^{19}$ erg/s$)(1.4 \times 10^{17}$ s$)/M_S \sim 3 \times 10^{10}$ erg/g is comparable to both the latent heat of vaporization of water, ~ $2 \times 10^{10}$ erg/g, and to the specific kinetic energy needed to escape Europa's gravity, which is also ~ $2 \times 10^{10}$ erg/g. If Europa was initially ice-rich, Io could have been as well by the above arguments, given its much higher observed heat flow, $F_t \sim 10^{21}$ erg/sec (*e.g.*, *Moore et al.*, 2007).

It is not clear, however, that conversion of tidal heat into vaporization and water vapor loss could have been efficient in an early, ice-rich Europa. Once its rocky core was surrounded by a thick layer of water and an ice shell, transport of hot material to the surface would likely have been limited (McKinnon, personal communication). If conversion of tidal heat into water vaporization was substantially less than 100% efficient at Europa, Europa's ice deficit is likely primordial, supporting the standard view that it formed predominantly in an environment too warm for ice. In this case Io would have also formed without substantial ice, because by virtue of being interior to Europa it would have been subject to higher disk temperatures.

## 2.4 Callisto's interior

*Galileo* gravity data allow for estimates of the satellites' axial moments of inertia, *C* (Table 1), which can be used to constrain the degree of differentiation under the assumption that the satellites are in hydrostatic equilibrium. Such models imply that Io, Europa, and Ganymede are differentiated. For example, three-layer interior models for Ganymede that can account for its bulk density and low $C/M_S R_S^2 = 0.3115 \pm 0.0028$ value (where $M_S$ and $R_S$ are the satellite's mass and radius) have a metallic core overlain by a rock mantle and an ice shell (*e.g.*, *Schubert et al.*, 2004). It is unknown whether differentiation in the inner three Galilean satellites occurred during their accretion or long after, with the latter possible due to a combination of tidal heating, resonance passage(s) associated with their evolution into the Laplace resonance, and/or radiogenic heating from their rock components (*e.g.*, *Friedson and Stevenson*, 1983; *Malhotra*, 1991; *Showman et al.*, 1997).

Unlike the inner Galilean satellites, outermost Callisto appears to be largely undifferentiated. Callisto's $C/M_S R_S^2 = 0.3549 \pm 0.0042$ (*Anderson et al.*, 2001) is higher than that of differentiated Ganymede but lower than that predicted for a completely undifferentiated state ($C/M_S R_S^2 \sim 0.38$ including compression effects, *McKinnon*, 1997). An example two-layer model for Callisto (*Anderson et al.*, 2001) invokes an outer ice shell ~300 km thick overlying an undifferentiated, ~ 2100 km mixed rock-ice interior. While such solutions are non-unique, all involve some undifferentiated component in order to account for Callisto's inferred $C/M_S R_S^2$ (*Schubert et al.,* 2004). Some uncertainty in this interpretation persists, because the *Galileo* flybys were nearly equatorial and because Callisto rotates slowly. As a result, available data do not independently constrain $J_2$ and $C_{22}$, and so it has not been possible to determine the ($J_2/C_{22}$) ratio as an independent check on whether Callisto is well described by a hydrostatic state. If Callisto is non-hydrostatic, it is possible that it could be differentiated and still have a $C/M_S R_S^2$ consistent with the limited *Galileo* data, for example due to structure on the surface of its rocky core (*e.g.*, *McKinnon*, 1997). However, hydrostatic equilibrium is generally believed to be a good assumption given Callisto's size (*Schubert et al.,* 2004).

At face value, *Galileo* data imply that Callisto is only partially differentiated. Because differentiation is an irreversible event, the presence of a mixed ice/rock core or mantle in Callisto requires that the satellite avoided wide scale melting of its ice throughout its entire history, including during its formation. This requirement in turn provides constraints on the rate and timing of Callisto's accretion, and by extension on the accretion rate and formation time of the other satellites as well.

The gravitational binding energy per unit mass associated with assembling Callisto is ~ $(3/5)GM_S/R_S$ ~ $1.8 \times 10^{10}$ erg/g, which is much larger than the latent heat of fusion for water ice, $l$ ~ $3 \times 10^9$ erg/g. If a Callisto-sized satellite formed rapidly in $\leq 10^4$ yr, its ice would melt throughout nearly its entire volume (*e.g.*, *Stevenson et al.*, 1986), leading to complete differentiation (*Friedson and Stevenson*, 1983). Avoiding this outcome requires that Callisto formed slowly, so that radiative cooling counteracts accretional heating. The coldest satellite is achieved in the limit that Callisto forms by the accretion of small, low-velocity objects that deposit their impact energy at or very near its surface where it can be radiated away. In this limit, the temperature $T_s$ at Callisto's surface is determined by the difference between the rates of accretional energy delivery and radiative cooling (*e.g.*, *Pritchard and Stevenson*, 2000; *Barr and Canup*, 2008):

$$\frac{1}{2}\dot{M}_S v_{imp}^2 - \sigma_{SB}(T_s^4 - T^4)4\pi R_S^2 = \dot{M}_S C_p (T_s - T_0), \qquad (2.1)$$

where $M_S$ and $R_S$ are the time-dependent satellite mass and radius, $v_{imp}$ is the characteristic impact velocity of accreting material, $\sigma_{SB}$ is the Stefan-Boltzmann constant, $T_0$ is the initial temperature of accreting matter, $T$ is the temperature in the circumplanetary disk, and $C_p$ ~ $1.5 \times 10^7$ erg/g/K is specific heat. Equation (2.1) assumes that the satellite radiates into an optically thick disk, which is probable for standard dust opacities (see § 4.1.2). In the region where Callisto formed, $T$ was likely ~ 100 K (*e.g.*, § 4.1.2 and Fig. 5b), while the minimum impact velocity is the satellite's escape velocity, $v_{esc} \equiv (2GM_S/R_S)^{1/2}$. Defining the accretion timescale $\tau_{acc} \equiv M_S / \dot{M}_S$, setting $v_{imp} = v_{esc}$ and $T = T_0 = 100$ K, and solving for the minimum $\tau_{acc}$ necessary to avoid melting of water ice in Callisto gives $\tau_{acc} \geq 5 \times 10^5$ yr (*Barr and Canup*, 2008; *cf. Stevenson et al.*, 1986)

Another potential heat source to a growing Callisto is radiogenic heating due to the decay of short-lived radioisotopes in the satellite's rocky components. The amount of radiogenic heating depends strongly on the absolute time that the satellite forms, because the primary heat source, $^{26}$Al, decays with a half-life of 0.716 Myr. For Callisto to avoid melting due to radiogenic heating alone (assuming no retention of accretional energy), it must have completed its accretion no earlier than ~ 2.6 to 3 Myr after the origin of the calcium-aluminum rich inclusions, or CAI's (*McKinnon*, 2006). Including both accretional and radiogenic heating implies that Callisto's formation was completed no earlier than about 4 Myr after CAI's for a disk temperature $T$ ~ 100 K (*Barr and Canup*, 2008).

An incompletely differentiated Callisto thus provides constraints on the start time of accretion, its accretion timescale, and therefore on the end time of its accretion. To avoid melting, the satellite must have *i*) started accreting at least several million years after CAI's, *ii*) accreted its mass slowly, in ~ $5 \times 10^5$ yr or more, and *iii*) completed its accretion no earlier then ~ 4 Myr after CAI's, given the total time implied by (*i*) and (*ii*). The completion time in (*iii*) is similar to mean observed circumsolar disk lifetimes (~ 3 Myr; *Haisch et al.*, 2001) and the time needed to grow Jupiter based on current core accretion models (~ 5 Myr; *Hubickyj et al.*, 2005). Requirement (*ii*) that $\tau_{acc}$ must be $\geq 5 \times 10^5$ yr provides the most restrictive constraint on disk conditions, in particular on the supply rate of material to the disk. If all the material needed to form Callisto were orbiting Jupiter near the satellite's current orbit (the presumed starting condition of an MMSN disk), the satellite would accrete in ~ $10^3$ yr (*e.g.*, *Canup and Ward*, 2002) because of Callisto's short orbital period of only ~ 2 weeks. Yet Callisto must have formed orders-of-magnitude more slowly to avoid melting. Assuming that the Galilean satellites formed together through common overall processes, similarly long formation times are then implied for Io, Europa and Ganymede as well.

### 2.5 Summary

Constraints that must be satisfied by any Galilean satellite origin model thus include 1) creation of a protosatellite disk to which a minimum of ~ $2 \times 10^{-4} M_J$ in solids is supplied (corresponding to ~ $0.02 M_J$ in solar composition material), 2) final satellites with similar individual masses comparable to that of the Earth's Moon, 3) an outer edge to the regular satellite region of approximately a few tens of Jovian radii, 4) temperatures low enough for incorporation of substantial water ice into Ganymede and Callisto, and 5) protracted and relatively late satellite accretion, assuming that Callisto is indeed only partially differentiated. We now turn to a discussion of how such constraints might naturally be satisfied during the late stages of gas accretion onto Jupiter.

## 3.    CIRCUMPLANETARY DISK DEVELOPMENT

As discussed in the introduction, a gas planet can develop a disk when either: (*i*) its rotation rate increases to the point of rotational instability, so that the planet starts to shed material creating a so-called spin-out disk, or (*ii*) the radius of the planet decreases below the outer radial extent of the pattern of inflowing gas, so that material flows directly into an accretion disk rather than directly onto the planet. Whether a spin-out or an accretion disk predominated for Jupiter is a function of the planet's gas accretion and contraction history.

Contraction models often assume that Jupiter accreted all of its mass prior to its contraction (*e.g., Bodenheimer and Pollack*, 1986). However, it appears likely that Jupiter would have still been accreting gas while it contracted. *Papaloizou and Nelson* (2005) tracked a Jovian planet's evolution including ongoing gas accretion, modeling the planet's state by numerically integrating stellar structure equations modified by the existence of a solid core (an approach also used by, *e.g., Bodenheimer and Pollack*, 1986, and *Pollack et al.*, 1996). Two classes of models were considered: Model A assumed the planet extends to the boundary of the Hill sphere and is relevant to the early phase of accretion, while Model B assumed a free boundary of a contracting planet that accretes mass from a circumplanetary disk, appropriate for Jupiter's final growth stages. The *Papaloizou and Nelson* (2005) models do not track the accreted angular momentum. Yet the existence of a circumplanetary disk at all implies a net angular momentum content of the inflowing gas, and for self-consistency, one must account for the removal of any excess angular momentum that would otherwise inhibit the planet's contraction. *Ward and Canup* (in preparation) have developed a model of disk formation around an accreting and contracting gas planet that focuses in particular on the angular momentum budget and the circumplanetary disk properties.

In this section we begin by reviewing estimates of the net angular momentum content expected for gas inflowing to Jupiter and the implied point at which rotational instability will occur as the planet contracts. We use this to first consider the simplest case of the contraction of a full-mass Jupiter in the limit of no gas accretion and no gas viscosity, and estimate the properties of a resulting spin-out disk. We then consider what we believe to be the more realistic case of a Jupiter that accretes gas as it contracts and is accompanied by a viscous circumplanetary disk. In this case, although a spin-out disk forms during the earliest stages of the planet's contraction, an accretion disk dominates once the planet has contracted to within the region of the current satellites.

### 3.1 Angular Momentum of Inflowing Gas

The net inflow of gas to Jupiter contained some average specific angular momentum, $j_c(t)$. This is difficult to estimate precisely without detailed hydrodynamic modeling, but some dimensional arguments can be employed to provide rough estimates (*e.g., Lissauer and Kary*, 1991). The planet will accrete gas parcels with both positive and negative angular momentum with respect to its center of mass. If we assume that the planet accretes all material that passes within some distance $R'$ from its center, and that the gas flows on approximately Keplerian orbits, the net angular momentum delivery rate to the planet, $\dot{L} \sim \ell \dot{M} R'^2 \Omega_J$, can be simply obtained by integrating over the accreting region. Here $\dot{M}$ and $\Omega_J$ are the planet's mass accretion rate and orbital frequency (so that $\Omega_J \equiv \sqrt{GM_\odot / a_J^3}$ for Jupiter), and $\ell$ is the angular momentum bias, with $\ell > 0$ corresponding to a prograde rotating planet. For a two-dimensional inflow, appropriate if gas is accreted primarily within the planet's orbital plane, $\ell \approx 1/4$ (*Ruskol*, 1982; *Lissauer and Kary*, 1991; *Mosquiera and Estrada*, 2003). If we set the radius of the accreting region, $R'$, to be comparable to the planet's Hill radius, $R_H \equiv a_J (M / 3M_\odot)^{1/3}$, then

$$j_c \approx \ell R_H^2 \Omega_J = \frac{a_J^2 \Omega_J}{4} \left( \frac{M}{3M_\odot} \right)^{2/3}, \qquad (3.1)$$

where $M$ is the planet's time-dependent mass. Other rationales for choosing $R'$ are discussed by *Ward and Canup* (in preparation), but for simplicity we use (3.1) in the discussion here. We define a centrifugal radius for the inflowing material, $r_c$, as the distance from the planet's center at which the specific angular momentum of a circular orbit, $(GMr_c)^{1/2}$, equals that of the inflow average, $j_c$ (*e.g., Cassen and Summers*, 1983). For prescription (3.1), $r_c = \ell^2 R_H/3 \sim 0.02 R_H$, so that inflow to a full-mass Jupiter would have a centrifugal radius $r_c \sim 15 R_J$.

Of course it is unlikely that at any given time, all of the inflowing material has a specific angular momentum, $j$, equal to that of the average, $j_c$. Hydrodynamical simulations have not yet determined this angular momentum distribution. In the absence of such information, we adopt a heuristic model in which incoming mass $\Delta M$ is distributed evenly across a range of specific angular momenta $(j_c - \Delta j/2) < j < (j_c + \Delta j/2)$, so that there is a mass $dM \sim (\Delta M /\Delta j)\, dj$ in any interval $dj$. When this material achieves centrifugal balance at an orbital distance $r$ from the planet's center, the corresponding range of specific angular momenta is $dj = (GMr)^{1/2}\, dr/2r$. For a mass inflow rate $\mathcal{F}$, this implies an inflow rate per unit area, $f_{inflow}$, to the planet's and/or disk's mid-plane of

$$f_{inflow} = \frac{\mathcal{F}}{4\pi r_c^2}\left(\frac{j_c}{\Delta j}\right)\left(\frac{r_c}{r}\right)^{3/2}. \qquad (3.2)$$

Here we will adopt $\Delta j \sim j_c$ for specificity, so that the inner and outer radial boundaries of the inflow region are $r_i \sim r_c /4$ and $r_o \sim 9 r_c /4$. For a full-mass Jupiter, this corresponds to a region extending from about $4 R_J$ to $34 R_J$. Although based on a simplified estimate of the angular momentum content of the inflowing gas, this range is roughly comparable to the Galilean satellite region. For comparison, three-dimensional hydrodynamic simulations with high spatial resolution within the planet's Hill sphere find an average specific angular momentum for gas inflowing to a Jupiter mass planet of $j_c \sim 0.3 R_H^2 \Omega_J$, or $r_c \sim 22 R_J$ (*Machida et al.*, 2008; *Machida*, 2008).

### 3.2 Rotational Instability

As Jupiter contracts, its spin frequency, $w \sim L/C$, increases, where $L$ is its total angular momentum, $C = l M R^2$ is its moment of inertia, and we will assume for simplicity that convection maintains nearly uniform rotation within the planet. A moment of inertia parameter can be roughly estimated from a polytropic model (*e.g.*, *Chandrasekhar*, 1958). For example, for a polytropic index $n = 2.5$, appropriate for molecular hydrogen and the early stages of Jupiter's contraction, $l \approx 0.17$. For comparison, Jupiter's current value is $l = 0.26$.

When the planet's spin frequency reaches the critical value for rotational instability, $\omega_c \equiv (GM / R^3)^{1/2}$, the planet starts to shed material from its equator into the disk. The planet's cumulative angular momentum, $L$, is found by integrating (3.1), with $L =$ $\int j_c dM = (3\ell/5)MR_H^2\Omega_J$. For a full mass Jupiter, this gives $L = \ell(5.4 \times 10^{47})$ g·cm$^2$/sec (much larger than the current angular momentum of Jupiter and its satellites augmented to solar composition, $\sim 8.6 \times 10^{45}$ g·cm$^2$/sec; *e.g.*, *Korycansky et al.*, 1991). As Jupiter contracts, rotational instability sets in once $L$ equals the critical value, $L_c = \mathbf{1}MR^2\mathbf{w}_c$, which determines the critical planet radius, $R_{rot}$, for commencement of spin-out from the planet,

$$R_{rot} = \frac{1}{3}\left(\frac{3\ell}{5\lambda}\right)^2 R_H \quad . \tag{3.3}$$

With $\ell = 1/4$, $\lambda = 0.17$, and $R_H = 744R_J$, this gives $R_{rot} \sim 200R_J$.

### 3.3 Contraction of a Full-Mass Jupiter: Properties of an Inviscid Spin-Out Disk

We first examine a case in which the bulk of the planet's mass has already been accreted before rotational instability sets in, so that there is no additional mass inflow to the planet as it contracts (*i.e.*, $\mathscr{F} \approx 0$). Once $R \leq R_{rot}$, the planet begins to shed mass into a disk.

The behavior of the disk material depends on its ability to transport angular momentum via viscous stresses. If we consider the extreme of a completely inviscid disk, shed material remains in orbit where it first leaves the planet. In the absence of inflow, the change in the planet's mass is equal to that shed to the disk, $dM = 2\pi r\sigma_g dR$, where $\sigma_g$ is the disk surface density. As the planet sheds mass at rate $\dot{M}$, its angular momentum decreases at a rate $\dot{L} = \dot{M}(GMR)^{1/2}$. Requiring the planet to maintain a critical rotation rate with $L = L_c = \lambda M(GMR)^{1/2}$ gives

$$\dot{L} = L\left(\frac{3}{2}\frac{\dot{M}}{M} + \frac{1}{2}\frac{\dot{R}}{R}\right), \tag{3.4}$$

which together with $\dot{L} = \dot{M}(GMR)^{1/2}$ implies

$$\frac{R}{M}\frac{dM}{dR} = \frac{\lambda}{2-3\lambda} \equiv q \quad . \tag{3.5}$$

This is easily integrated to find the mass retained in the contracting planet as a function of its radius, $M = M_0(R/R_{rot})^q$, where $q = 0.11$ for $\mathbf{1} = 0.17$, and $M_0$ is the planet's initial mass before its contraction. The total mass deposited in the spin-out disk is then $M_d = M_0[1-(R/R_{rot})^q]$, with the final disk mass obtained by setting $R \approx R_J << R_{rot}$. If $R_{rot} \approx 10^2 R_J$, $M_{final} \sim 0.6M_0$ so that requiring $M_{final} = M_J$ implies $M_0 = 1.66\ M_J$. The resulting inviscid spin-out disk would then contain $\sim 0.4M_0$, or $\sim 66\%$ of a Jupiter mass. If $R_{rot}$ were $\approx 25R_J$, due, for example, to a factor of 2 lower value for $\ell$, the disk mass decreases somewhat but is still $\sim 0.4M_J$. These disks are vastly more massive than the MMSN disk, which contains $\sim 0.02M_J$. In addition, such massive disks would likely be gravitationally unstable, which would generate a viscosity and invalidate the starting assumption of an inviscid disk.

The discussion above assumes a largely convective and uniformly rotating planet. An alternative model is described by *Korycansky et al.* (1991), in which the planet's structure includes a differentially rotating radiative region of constant specific angular momentum between two uniformly rotating convective regions. This results in a more centrally condensed object with a larger critical radius for rotational instability. In this case, a smaller total disk mass is shed, but the disk has a much larger outer edge, $\sim 300$ planetary radii for the case of a Saturn-mass planet.

Thus given the estimated angular momentum budget associated with accreting Jupiter, the angular momentum that would need to be contained in an inviscid spin-out disk is so large that the resulting disk is too massive and/or too radially extended compared to the MMSN construct. To produce an appropriate MMSN-type disk would require instead that $\ell \sim 0.02$, or that $R'$ be significantly less than the planet's Hill radius, which seems difficult to justify given results of hydrodynamic simulations (*e.g.*, Figure 1).

## 3.4 Viscosity in the Circumplanetary Disk

The starting condition of a static, non-viscous MMSN disk invoked in much early work on satellite formation appears difficult to reconcile with current angular momentum estimates and the subsequent contraction process for a Jupiter-like planet. However, a number of factors suggest that once gas is in circumplanetary orbit, it will not remain static but will instead radially spread due to viscous angular momentum transport. Perhaps the most compelling argument for viscosity is that viscous diffusion allows most of the disk's mass to evolve inward and be accreted by the planet, while most of the disk's angular momentum (contained in a small fraction of its mass) is transported outwards where it can eventually be removed from circumplanetary orbit. Thus a much smaller disk mass can be consistent with the expected angular momentum budget than in the inviscid case. A circumplanetary disk also shares basic traits with the circumsolar disk, in which viscosity is commonly invoked to explain, *e.g.*, observed mass accretion rates onto stars in extrasolar systems (*e.g.*, *Stone et al.*, 2000).

The source and magnitude of viscosity in cold disks is actively debated, and while a number of candidate mechanisms have been proposed (see § 5), the mode and rate of angular momentum transport are quite uncertain. Given this uncertainty, a simplified parameterization is often used. The so-called "alpha model" (*Shakura and Sunyaev*, 1973) defines a viscosity $\nu = \alpha c H \approx \alpha c^2 / \Omega$, where $\alpha$ is a constant, $c$ is the gas sound speed in the disk, and $H \approx c/\Omega$ its vertical scale height (*i.e.*, the disk's half thickness), with $\Omega \equiv (GM / r^3)^{1/2}$ the orbital frequency at radius $r$ in the disk. The alpha model envisions angular momentum transport due to communication between turbulent disk eddies. The $\alpha$-viscosity expression is similar to that for molecular viscosity, only with the molecular velocity replaced by an eddy velocity, $\alpha^{1/2} c$, and the molecular mean free path replaced by an eddy length scale, $\alpha^{1/2} H$ (*e.g.*, *Dubrulle et al.*, 1995). Because the characteristic eddy velocity and length scale would be no larger than the disk sound speed and the disk thickness, respectively, the $\alpha$ parameter is a constant less than unity. Circumstellar disk models typically consider $10^{-4} \leq \alpha \leq 0.1$ (*e.g.*, *Stone et al.*, 2000).

Angular momentum transport is characterized by a torque due to viscous shear, $g = 3\pi \sigma_g \nu j$, exerted by the inner disk on the outer disk across the circumference $2\pi r$, where $\sigma_g$ is the gas surface density, and $j = (GMr)^{1/2}$ is the specific angular momentum at orbital radius $r$; $g$ is often referred to as the viscous couple (*Lynden-Bell and Pringle*, 1974). The viscous torque causes the angular momentum of an annular ring of mass $\delta m = 2\pi r \sigma_g \delta r$ to change at a rate

$$\delta m \frac{d}{dt}(j) = g(r) - g(r + \delta r) = -\frac{\partial g}{\partial r} \delta r \qquad (3.6)$$

With $(d/dt)j = (\partial j / \partial r)u$ , where $u = (dr/dt)$ is the radial velocity of disk material with $u > 0$ defined as an outward velocity, equation (3.6) gives

$$2\pi r \sigma_g \delta r \frac{\partial j}{\partial r} u = -\frac{\partial g}{\partial r} \delta r \qquad (3.7)$$

Thus the couple drives an in-plane radial mass flux in the disk, $F = 2\pi r \sigma_g u$ , and $F$ satisfies (*Lynden-Bell and Pringle*, 1974)

$$F\frac{\partial j}{\partial r} = -\frac{\partial g}{\partial r} \quad . \tag{3.8}$$

When the outer edge of the disk expands to a distance comparable to the planet's Hill radius, material escapes from the gravity of the planet and is returned to circumstellar orbit. Accordingly a viscous circumplanetary disk will have an outer edge $r_d = \gamma R_H$, where $\gamma$ is a moderate fraction of unity, and a vanishing couple there, with $g(r_d) = 0$. Provided the viscous spreading time, $\tau_\nu \sim r_d^2 / \nu$, is less than the characteristic time scale over which processes supplying material to the disk change, a quasi-steady state is attained wherein the time-variation in the flux $F$ can be ignored when integrating (3.8), so that $F \approx F(r)$ (see also section 3.5.5).

## 3.5 Contraction of Jupiter with Ongoing Gas Inflow: Evolution from a Viscous Spin-Out to an Accretion Disk

In § 3.2 we considered a planet that had acquired its full mass prior to its contraction. However, estimates of maximum gas accretion rates ($\sim 10^{-2}$ $M_\oplus$/yr, *e.g. Hayashi et al.*, 1985) imply that Jupiter would begin to contract to a size smaller than its Hill sphere once it exceeded $\sim 0.2$ to $0.3 M_J$ (*e.g.*, *Tajima and Nakagawa*, 1997). In addition, predicted contraction timescales for Jupiter to achieve a size comparable to or less than the regular satellite region are shorter than the expected nebular lifetime and the duration of the inflow (*e.g. Papaloizou and Nelson*, 2005). Thus it seems most likely that for Jupiter, contraction and accretion occurred simultaneously. In this case, the system passes through stages in which the circumplanetary disk transitions from a spin-out disk to an accretion disk, as we next describe.

A comparison of the outer radial boundary for inflowing gas, $r_o \sim (1/3)(3\ell/2)^2 R_H$ (§ 3.1), with the critical radius for instability $R_{rot}$ from equation (3.3), reveals that $R_{rot}/r_o \sim (2/5\mathit{l})^2 \approx 5.5$ for $\mathit{l} = 0.17$, with $(R_{rot}/r_o)$ independent of the planet's mass. Thus for nominal parameter choices, Jupiter will begin to shed material into a spin-out disk first, while it is still larger than the outer boundary of the effective pattern of inflowing gas.

Figure 3 illustrates the three stages that Jupiter and its disk pass through as a function of the planet's time-dependent radius, $R$. A disk first appears when $r_o < R < R_{rot}$ (stage 1). During stage 1, inflowing material falls directly onto the planet, but because the planet is rotating at the critical rate it must shed material to the disk as it contracts. Thus stage 1 involves a purely spin-out disk. As the planet contracts further so that $r_i < R < r_o$ (stage 2), part of the inflowing gas falls directly onto the planet while part falls onto the disk, and the resulting disk can either be predominately a spin-out or an accretion disk. By the time Jupiter contracts to a radius smaller than that of the inner edge of the inflow pattern ($R < r_i$, stage 3), all of the inflowing gas falls onto the disk, which approaches a pure accretion disk. We now discuss the properties of the circumplanetary disk as a function of time and position within the disk, assuming that the disk maintains a quasi-steady state, following the framework of *Ward and Canup* (in preparation). At the end of this section we discuss when the steady state assumption is likely to be valid.

### 3.5.1 *In-plane mass flux.*

Wherever inflow impacts the disk, the continuity equation gives (*Canup and Ward*, 2002),

$$\frac{\partial \sigma_g}{\partial t} = f_{inflow} - \frac{1}{2\pi r}\frac{dF}{dr} \ ,$$  (3.9)

which for a disk in steady-state (*i.e.*, $(\partial \sigma_g / \partial t) = 0$) implies

$$\frac{dF}{dr} = 2\pi r f_{inflow} = \frac{\mathscr{F}}{2 r_c^{1/2} r^{1/2}} \ ,$$  (3.10)

where equation (3.2) has been used to relate the inflow rate per disk area, $f_{inflow}$, to the global mass inflow rate, $\mathscr{F}$. Note that while ($dF/dr$) is positive definite, $F$ itself can be positive or negative, depending on the position within the disk. Integration of (3.10) gives the flux variation with $r$ for the various radial zones in the disk. We define two distance markers: the first, $r_s$, changes with the stage of the planet's contraction, while the second, $r_z$, changes with the position (zone) within the disk, *viz.*,

$$r_s(R) \equiv \begin{cases} r_o & \text{for } r_o < R \text{ (stage 1)} \\ R & \text{for } r_i < R < r_o \text{ (stage 2)} \\ r_i & \text{for } R < r_i \text{ (stage 3)} \end{cases} \quad \text{and} \quad r_z(r) \equiv \begin{cases} r_o & \text{for } r_o < r \text{ (zone 1)} \\ r & \text{for } r_i < r < r_o \text{ (zone 2)} \\ r_i & \text{for } r < r_i \text{ (zone 3)} \end{cases}$$  (3.11)

Using these, the flux can then be written compactly as

$$F(r) = F_p + \mathscr{F}\left[\left(\frac{r_z}{r_c}\right)^{1/2} - \left(\frac{r_s}{r_c}\right)^{1/2}\right] \ ,$$  (3.12)

where $F_p$ is the flux at the inner edge of the disk denoting the mass exchange rate between the disk and the planet. Evaluating (3.12) at $r_d$ tells us that the flux exiting the disk at the outer edge is $F_d = F_p + \mathscr{F}\,[(r_o/r_c)^{1/2} - (r_s/r_c)^{1/2}]$. With this, a useful alternative expression for $F$ in terms of $F_d$ can be found

$$F(r) = F_d - \mathscr{F}\left[\left(\frac{r_o}{r_c}\right)^{1/2} - \left(\frac{r_z}{r_c}\right)^{1/2}\right] \tag{3.13}$$

### 3.5.2 *The viscous couple.*

To determine the structure of the disk, we integrate (3.8) with the appropriate boundary conditions for each zone. In zones 1 and 3 where the flux is constant in $r$, the combination $Fj + g$ will be constant as well. However, in zone 2 where there is inflow onto the disk, we use $Fdj/dr = d(Fj)dr - j\,dF/dr$ to rewrite (3.8) as

$$\frac{d}{dr}(Fj + g) = j\frac{dF}{dr} \tag{3.14}$$

and integrate with the help of (3.10) (*Canup and Ward*, 2002). This gives

$$g = g_p + F_p j_p - Fj + \frac{\mathscr{F}}{2}\left[\left(\frac{r_z}{r_c}\right) - \left(\frac{r_s}{r_c}\right)\right]j_c \ , \tag{3.15}$$

where $j_p = j(R)$, $j_c = j(r_c)$, and $g_p = g(R)$. Next, the boundary condition that the couple vanishes at the outer edge is applied to constrain its value at the planet, *i.e.*,

$$g_p = F_d j_d - F_p j_p - \frac{\mathscr{F}}{2}\left[\left(\frac{r_o}{r_c}\right) - \left(\frac{r_s}{r_c}\right)\right]j_c \ , \tag{3.16}$$

where $j_d = j(r_d)$. Equations (3.13) and (3.16) can now used to eliminate $F$ and $g_p$ in (3.15) to yield (*Ward and Canup*, in preparation),

$$g = F_d(j_d - j) + \mathscr{F}\left[\left(\frac{r_o}{r_c}\right)^{1/2} - \left(\frac{r_z}{r_c}\right)^{1/2}\right]j \ - \frac{\mathscr{F}}{2}\left[\frac{r_o}{r_c} - \frac{r_z}{r_c}\right]j_c \ . \tag{3.17}$$

In a quasi-steady state, conservation of mass dictates that the rate of inflow to the system must equal the rate of change of the planet's mass plus any mass loss at the outer edge, *i.e.*, $\mathscr{F} = \dot{M} + F_d$. Substituting this into (3.17), setting $r_o = 9r_c/4$, $g = 3\pi\sigma_g\nu j$, and dividing throughout by $j$ gives

$$3\pi\sigma_g\nu = (\mathscr{F} - \dot{M})\left[\left(\frac{r_d}{r}\right)^{1/2} - 1\right] + \mathscr{F}\left[\frac{3}{2} - \left(\frac{r_z}{r_c}\right)^{1/2}\right] - \frac{\mathscr{F}}{2}\left[\frac{9}{4} - \frac{r_z}{r_c}\right]\left(\frac{r_c}{r}\right)^{1/2}. \quad (3.18)$$

The remaining unknown is $\dot{M}$, and to find this, the planet's evolution including the disk interaction must now be considered.

### 3.5.3 Evolution of the planet.

A constraint on $\dot{M}$ is provided by the planet's angular momentum evolution,

$$\dot{L} = \frac{\mathscr{F}}{2}\left[\frac{r_s}{r_c} - \frac{r_i}{r_c}\right]j_c - \left(F_p j_p + g_p\right) \quad (3.19)$$

where the first term is the angular momentum falling directly on the planet. Using (3.16) to eliminate $g_p$, this becomes $\dot{L} = \mathscr{F}j_c - F_d j_d = \mathscr{F}j_c + (\dot{M} - \mathscr{F})j_d$, which when equated to (3.4) implies,

$$\left[3 - \frac{2}{\lambda}\left(\frac{r_d}{R}\right)^{1/2}\right]\frac{\dot{M}}{M} + \frac{\dot{R}}{R} = -\frac{2}{\lambda}\left(\frac{r_d}{R}\right)^{1/2}\frac{\mathscr{F}}{M}\left[1 - \left(\frac{r_c}{r_d}\right)^{1/2}\right]. \quad (3.20)$$

In the limit $(r_d/R)^{1/2} \gg 1$, this expression reduces to

$$\dot{M} \approx -\frac{\lambda}{2}\left(\frac{R}{r_d}\right)^{1/2}\left|\frac{\dot{R}}{R}\right|M + \mathscr{F}\left[1 - \left(\frac{r_c}{r_d}\right)^{1/2}\right] \quad (3.21)$$

Note that the disk outflow is

$$F_d = \mathscr{F} - \dot{M} \approx \frac{\lambda}{2}\left(\frac{R}{r_d}\right)^{1/2}\frac{M}{\tau_{KH}} + \mathscr{F}\left(\frac{r_c}{r_d}\right)^{1/2}, \quad (3.22)$$

where we define a planet contraction timescale, $\tau_{KH} \equiv \left|R/\dot{R}\right|$. The flux across the disk's outer edge, $F_d$, is always positive and contains a contribution generated by the contraction and one by the inflow. However, the flux at the planet is

$$F_p = F_d - \mathscr{F}\left[\frac{3}{2} - \left(\frac{r_z}{r_c}\right)^{1/2}\right], \quad (3.23)$$

which can be either positive or negative. In stage 1, $F_p = F_d > 0$, and the disk is due to spin-out. However, $F_p$ progressively decreases from $F_d$ in stage 2, until in stage 3, $F_p = F_d - \mathscr{F}$, and the planet accretes material from the disk ($F_p < 0$) if

$$\frac{\lambda}{2}\left(\frac{R}{r_d}\right)^{1/2}\frac{M}{\tau_{KH}} < \mathscr{F}\left[1-\left(\frac{r_c}{r_d}\right)^{1/2}\right]. \tag{3.24}$$

### 3.5.4 *Example planet contraction and disk formation history.*

Equations (3.18) and (3.21) can be used to determine the evolution of the planet mass and the disk structure as a function of the inflow and contraction rates. Alternatively, if $\dot{M}(t)$ and $\dot{R}(t)$ are known from an accreting giant planet contraction model, such as given by *Papaloizou and Nelson* (2005, their Figure 11) shown here in Fig. 4a-b, equation (3.21) can be used to eliminate $\mathscr{F}$ in (3.18) Figure 4c shows the resulting behavior of $3\pi\sigma_g\nu$ as a function of $r$ at various times during the evolution. With time the inflow slows and the planet contracts, causing the inner disk edge to move inward and the peak value for $3\pi\sigma_g\nu$ to decrease. For the case shown here, once the planet contracts to a radius smaller than about $13R_J$, the flux at the planet becomes negative ($F_p < 0$) and the disk is increasingly well approximated as an accretion disk. Figure 4d shows the predicted gas surface density profile, $\sigma_g(r)$, at the time when the planet has accreted more than 90% of its final mass, assuming a simple model for the viscosity with $\nu \propto r^{1/2}$ (see Figure 4 caption). Shown for comparison is the analogous surface density profile using the simpler model of *Canup and Ward* (2002, see § 4), which considers a pure accretion disk and ignores possible contributions from spin-out. At late times in Jupiter's growth, the full disk model described in this section approaches that of a pure accretion disk.

### 3.5.5. *Discussion.*

The coupled planet-disk evolution model described here assumes that the viscous torque between the disk and the surface of the planet will maintain the planet's rotation rate at the limit of rotational stability as the planet contracts. Yet ultimately, Jupiter must be left with less angular momentum than this to account for its current ~ 10 hour rotational day, which is about a factor of 3 longer than the critical rotation period for an $M = M_J$ and $R = R_J$ planet.

Accounting for sub-critical rotation is a well-known issue for Jupiter and Saturn, as well as for protostars. Protostars, also believed to grow through mass delivered through a viscous accretion disk, have observed rotation rates that are often much slower than breakup (*e.g.*, *Herbst et al.*, 2002). Proposed solutions to account for protostar angular momentum loss may also apply to gas giant planets, including 1) "disk-locking", or magnetic coupling between the central object and its disk that results in angular momentum transferred to the disk beyond the co-rotation radius (where the Kepler velocity equals the primary rotational velocity; e.g., *Konigl*, 1991; *Takata and Stevenson*, 1996) and 2) magnetocentrifugally driven "X-winds" that originate from the inner accretion disk, diverting mass and angular momentum that would otherwise be delivered to the primary (*e.g.*, *Shu et al.*, 2000).

The first alternative has been explored in the context of a circumjovian disk by *Takata and Stevenson* (1996). From their work, one can derive a magnetic torque density of the form $dT_M / dr = -2B_J^2 R^2 \operatorname{Re}(R/r)^3 (\Omega - w)/w_c$, where $B_J$ is the mid-plane magnetic field strength at $R$, and Re is the magnetic Reynolds number that depends in part on the conductivity of the disk material. Integrating the torque density throughout the disk gives $T_M = B_J^2 R^3 \operatorname{Re}(w/w_c - 4/7)$. The magnetic torque on the planet, $-T_M$, is thus negative for $\omega > 4\omega_c / 7$, which could cause its rotation to become subcritical at some point as the inflow and disk mass decrease. The current spin rate of Jupiter is about 0.3 times its critical value. However, if the disk surface density becomes low enough that the magnetic torque density can force it to corotate with the planet (as a magnetosphere) out to some distance $r_M > R$, the total magnetic torque on the planet can remain negative for rotation rates below $4w_c / 7$. A full treatment of this issue will involve modifying the disk profile to include the magnetic torque and assessing the expected degree of ionization in a gas-starved disk, which will be a topic of future work.

The model described in this section also assumes that the in-plane flux can be approximated as a quasi-steady state. This is a valid approximation when the viscous spreading time of the circumplanetary disk, $\tau_v$, is short compared to both the planet's contraction timescale and the timescale over which the inflow changes. The viscous timescale is $\tau_v = r_d^2 / \nu \approx 3 \times 10^4$ yr $(10^{-3}/\alpha)(r_d / 300 R_J)^{3/2} (0.1/\{H/r\})^2$. In the planet's early contraction phase, the contraction timescale $\tau_{KH}$ was likely shorter than $\tau_v$ (*e.g.*, Fig. 4b), so that the resulting disk would be more massive than that predicted with a steady state analysis, although this may be mitigated by gravitational instability which could increase the effective viscosity. However, once the planet's contraction rate slows so that $\tau_v < \tau_{KH}$, disk material left in orbit earlier has time to viscously spread out to $r_d$, and a quasi-steady state is achieved. Similarly, an early, potentially rapidly varying gas inflow to the planet would produce a non-steady state disk, if a disk existed at all at that time. However during the planet's final growth, changes in the gas inflow rate were most likely regulated by the removal of the solar nebula itself with a characteristic timescale of $\sim 10^6$ to $10^7$ yr (*e.g.*, *Haisch et al*., 2001), which is much longer than $\tau_v$. Thus as Jupiter approached its final mass, a surrounding viscous disk would have been well described as a quasi-steady state.

## 4. DISK CONDITIONS AND GROWTH OF SATELLITES

We argued in Section 3 that in the late stages of its formation, Jupiter was surrounded by an inflow-supplied accretion disk. This contrasts with models that consider the protosatellite disk as an isolated system with a fixed total mass and no ongoing gas or solid supply (*e.g.*, *Lunine and Stevenson*, 1982; *Mosqueira and Estrada*, 2003). An actively supplied accretion disk has been considered by multiple works. *Coradini et al*. (1989) computed steady state conditions within a viscous accretion disk, assuming that the instantaneous disk mass would need to be equivalent to that of the MMSN in order to produce the Galilean satellites. Creating a disk this massive required rapid inflow rates, which in turn implied high disk temperatures inconsistent with icy satellites. As such, *Coradini et al*. proposed a two-stage evolution, with an accretion disk forming during the first stage, followed by a second stage of satellite growth after gas inflow to the disk stopped and the disk cooled. *Makalkin et al*. (1999) observed that appropriately low disk temperatures for producing icy satellites could result during gas inflow if the inflow rate was slow, but found that in this case the disk mass at any one time would be much less than that of the MMSN. *Makalkin et al*. concluded that the total mass and temperature constraints needed to produce the Galilean satellites could not be simultaneously satisfied in an accretion disk. *Canup and Ward* (2001, 2002) showed that a disk produced by a slow-inflow would not only have low temperatures, but could also 1) allow for survival of Galilean-sized satellites against orbital decay due to density wave interactions with the gas disk, and 2) lead to slow satellite accretion, as needed for consistency with Callisto's apparent interior state. While the *Canup and Ward* disk density was much lower than the MMSN, they argued that an appropriate total mass in solids could nonetheless build up in the disk over time to form perhaps multiple Galilean-like systems, with earlier satellites lost to collision with the planet. *Stevenson* (2001a,b) also advocated an actively supplied circumjovian disk, which he referred to as a "gas-starved" disk. *Alibert et al.* (2005) coupled an accretion disk model to a time-dependent prescription for gas delivery to Jupiter, and predicted satellite compositional and loss patterns assuming a linear satellite mass growth rate. However, in their models, inflowing mass was delivered only at the outer edge of the circumplanetary disk, and angular momentum was not carried away by outflow from this point. Hence to maintain quasi-equilibrium, the assumed inflow would have to add almost no net angular momentum, and it is unclear why such a flow would form a disk. *Canup and Ward* (2006) performed direct simulations of satellite accretion within a disk supplied by an ongoing, time-dependent inflow and found that systems similar to the Galilean satellites could be created over a relatively broad range of disk properties.

In this section, we begin by evaluating conditions expected within an accretion disk at the time the Galilean satellites formed, including disk density and temperature profiles. We then discuss possible sources of solid material (rock and ice) to the disk, the conditions in which such material will tend to accumulate into satellites, and the expected satellite growth timescales. The effects of density wave interactions between the growing satellites and the background gas disk are then described, including the tendency for massive satellites to spiral into the planet through inward migration. We demonstrate that migration selects for a maximum satellite mass and a common total satellite system mass, and conclude by showing example satellite systems produced by *N*-body accretion simulations.

### 4.1 Properties of the Gas Disk

To characterize the protosatellite environment, we wish to estimate how the disk gas density and temperature evolve. In the discussion to follow, we adopt the disk model of *Canup and Ward* (2002), which assumes that Jupiter's disk was a pure accretion disk that derived its mass solely from gas inflowing from solar orbit. As we have shown above (*e.g.*, Fig. 4d), this should be a good approximation during the late stages of Jupiter's growth when spin-out from the planet can be neglected. However future work should incorporate models (such as that described in § 3) that treat the inflow and the planet's state self-consistently.

4.1.1 *Gas surface density.*

In a pure accretion disk, the gas surface density reflects a balance between the inflow supply of gas, and gas removal as it viscously spreads onto the planet or beyond the disk's outer edge, $r_d$ (*e.g.*, Figure 2). Using methods similar to those employed in § 3, the steady state gas surface density, $\sigma_g$, can be determined as a function of $\mathcal{F}$, $\nu$, $r_d$, and the outer edge of the inflow region, $r_o$ (*Lynden-Bell and Pringle*, 1974; *Canup and Ward*, 2002). For the simple case of a uniform inflow (*i.e.*, $f_{inflow}$ = constant for $r < r_0$), this gives

$$\sigma_g(r) \approx \frac{4\mathscr{F}}{15\pi\nu}\begin{cases} \dfrac{5}{4}-\left(\dfrac{r_o}{r_d}\right)^{1/2}-\dfrac{1}{4}\left(\dfrac{r}{r_o}\right)^2 \text{ for } r < r_o \\[2ex] \left(\dfrac{r_o}{r}\right)^{1/2}-\left(\dfrac{r_o}{r_d}\right)^{1/2} \text{ for } r \geq r_o \end{cases}, \qquad (4.1)$$

(*Canup and Ward*, 2002). The surface density is proportional to the mass inflow rate, $\mathscr{F}$, so a more massive circumplanetary disk is expected during rapid inflow and $\sigma_g$ decreases as the inflow wanes. The surface density is inversely proportional to $\nu$ because as the viscosity increases, the disk spreads more rapidly, which reduces $\sigma_g$ for a fixed inflow rate.

### 4.1.2 *Disk temperature.*

The disk temperature affects the composition of accreting material as well as the disk's viscosity in the alpha-model, in which $\nu$ depends on the square of the gas sound speed, $c^2$, and therefore on the disk mid-plane temperature, $T$, with $c^2 = \gamma\mathfrak{R}T/\mu_{mol}$ (where $\gamma = 1.4$ is the adiabatic index and $\mu_{mol} \sim 2$ g/mol for molecular hydrogen, and $\mathfrak{R} = 8.31 \times 10^7$ erg/(mol·K) is the gas constant). The disk is heated by luminosity from Jupiter (which contributes even for a radially optically thick disk if the disk's vertical scale height increases with distance from the planet), viscous dissipation, and energy dissipation associated with the difference between the free-fall energy of the incoming gas and that of a Keplerian orbit. These energy sources are balanced by radiative cooling, which occurs predominantly in the vertical direction for a disk that is thin compared to its radial extent.

Jupiter's luminosity provides an energy rate per unit area to the region between orbital radii $r$ to $(r+\Delta r)$,

$$\dot{E}_J \approx 2\sigma_{SB}T_J^4\left(\frac{R_J}{r}\right)^2 2\pi r\left[H(r+\Delta r)-H(r)\right]/2\pi r\Delta r \approx \frac{9}{4}\sigma_{SB}T_J^4\left(\frac{R_J}{r}\right)^2\left(\frac{c}{r\Omega}\right) \quad (4.2)$$

where $T_J$ is Jupiter's temperature, and a $T \propto 1/r^{3/4}$ dependence has been assumed in calculating $H(r)$ (*Canup and Ward*, 2002). The energy production rate per area due to viscous dissipation is $\dot{E}_\nu = (9/4)\nu\Omega^2\sigma_g$, while that due to the infalling gas is $\dot{E}_{in} \approx (GM_J/2r)(\mathscr{F}/\pi r_o^2)$. The equation for thermal balance is (*e.g.*, *Canup and Ward*, 2002)

$$\frac{9}{4}\left[\sigma_{SB}T_J^4\left(\frac{R_J}{r}\right)^2\frac{c}{r\Omega}+\zeta\alpha c^2\Omega\sigma_g\right] = 2\sigma_{SB}T_{eff}^4\left[1-(T_{neb}/T_{eff})^4\right] \qquad (4.3)$$

where $\zeta \equiv 1 + 3/2\left[(r_o/r)^2 - 1/5\right]^{-1}$ accounts for the $\dot{E}_{in}$ term (generally small compared to the other energy sources), $T_{neb}$ is the ambient temperature into which the circumplanetary disk radiates, $T_{eff}$ is the disk's effective temperature, and the factor of 2 on the right hand side is because the disk has both upper and lower surfaces. Equation (4.3) assumes a radiative disk; convective energy transport is generally a small contribution, even in a turbulent disk (*e.g., Ruden and Lin*, 1986).

Prior Galilean satellite formation models have set $T_{neb}$ equal to the solar nebula temperature at Jupiter's orbit, which recent works estimate to be ~ 100 K (*e.g., Garaud and Lin*, 2007). However a Jupiter-mass planet will likely open a gap in the circumsolar disk (*e.g., Bate et al.*, 2003), allowing the circumplanetary disk and planet to radiate to a reduced density and colder background over a substantial fraction of the available solid angle. Numerical simulations by *D'Angelo et al.* (2003) consider the thermohydrodynamic evolution of circumstellar disks containing Jupiter-sized protoplanets. Assuming standard dust opacities, they find optical depths within the gap region are typically of order unity (depending on the choice of viscosity), with gap temperatures of $T_{neb} \sim 20$ to $50$ K. Note, however, that because the term in equation (4.3) that accounts for the ambient temperature is proportional to $(T_{neb}/T_{eff})^4$, it becomes unimportant when the effective temperature substantially exceeds $T_{neb}$, which would be the case during most of the disk's evolution in the region of the regular satellites (Fig. 5a).

The temperature $T_{eff}$ is that of an equivalent blackbody emitting the same radiative flux as the disk. For an optically thick disk, $T_{eff}$ is the disk's surface temperature. For an optically thin disk, the disk is isothermal, but $T_{eff}$ is less than the disk temperature, reflecting a reduced emissivity. Given equation (4.3), an additional relationship is needed to determine the disk mid-plane temperature, $T$. A generalized calculation of a disk's vertical temperature profile is a complex radiative transfer problem, although analytical approximations exist for many cases. In the limit that viscous dissipation occurs primarily in the disk mid-plane, while heating due to Jupiter's luminosity and infalling mass are concentrated at the disk's surface, an approximate expression for the mid-plane temperature appropriate in both high and low optical depth regimes is (*Nakamoto and Nakagawa*, 1994):

$$2\sigma_{SB}T^4 = \left(1 + \frac{3}{8}\kappa_R\sigma_g + \frac{1}{2\kappa_P\sigma_g}\right)\dot{E}_v + \left(1 + \frac{1}{2\kappa_P\sigma_g}\right)(\dot{E}_J + \dot{E}_{in}) + 2\sigma_{SB}T_{neb}^4 \qquad (4.4)$$

where $\kappa_R$ and $\kappa_P$ are the disk's frequency-averaged Rosseland and Planck opacities. The Rosseland mean opacity is relevant for optically thick regions, while the Planck mean opacity is relevant for optically thin regions in which scattering can be ignored and cooling occurs predominantly in resonance lines of abundant species (*e.g.*, *Hubeny*, 1990). For a solar composition gas, $\kappa_P$ is always much larger than $\kappa_R$ at low temperatures; *e.g.*, at 500 K, $\kappa_P \sim O(10^{-1})$ cm²/g while $\kappa_R \sim O(10^{-5})$ cm²/g (*Helling et al.*, 2000). If dust grains are present, opacities are much higher, with $\kappa_R \sim \kappa_P \sim$ a few cm²/g (*e.g.*, *Semenov et al.*, 2003). In the next section, we will argue that small particles appear the most promising source of solid material to an accretion disk, and in this case it seems reasonable to imagine an accompanying population of dust grains, although their size distribution is unknown.

Eqs. (4.3) and (4.4) simplify when viscous dissipation is the dominant energy source (see also *Hubeny*, 1990), as is typically the case in most of the regular satellite region (*Canup and Ward*, 2002). For $T_{eff}^4 \gg T_{neb}^4$,

$$T_{eff}^4 \approx \frac{9\Omega^2}{8\sigma_{SB}}\nu\sigma_g = \frac{3\Omega^2 \mathscr{F}}{8\pi\sigma_{SB}}\left[1 - \frac{4}{5}\sqrt{\frac{r_o}{r_d}} - \frac{1}{5}\left(\frac{r}{r_o}\right)^2\right] \qquad (4.5)$$

$$T^4 \approx \left(1 + \frac{3}{8}\kappa_R\sigma_g + \frac{1}{2\kappa_P\sigma_g}\right)T_{eff}^4 \ . \qquad (4.6)$$

Equation (4.5) shows that the effective disk temperature is a function of the inflow rate, $\mathscr{F}$, but is *independent* of the gas surface density and viscosity. This is because from (4.1), $\sigma_g \propto 1/\nu$, so that the combination $(\nu\sigma_g)$ in (4.5) does not depend on viscosity. The behavior of the mid-plane temperature, $T$, depends on whether the disk is optically thick or thin. In an optically thick disk (with optical depth $\tau_R \equiv \kappa_R\sigma_g > 1$), the second term on the right hand side of (4.6) dominates. Because $\sigma_g \propto \mathscr{F}/\nu \propto \mathscr{F}/(\alpha T)$, this gives $T \propto (\mathscr{F}\kappa_R/\alpha)^{1/5}T_{eff}^{4/5} \propto \mathscr{F}^{2/5}(\kappa_R/\alpha)^{1/5}$, so that the mid-plane temperature depends rather weakly on the opacity and viscosity, and is also more strongly affected by the inflow rate. For a dusty disk with $\kappa_R \sim$ unity, the optically thick regime applies for $\sigma_g > O(1)$ g/cm², or for gas surface densities as small as $10^{-5}$ times that of the MMSN. For lower opacities, the optically thin regime applies for higher gas surface densities, and in this case the Planck term dominates, the disk is isothermal, and the viscosity and opacity are important in setting $T$.

For an optically thick disk supplied by an inflow, the inflow rate is a key control on the disk mid-plane temperature, unlike the frequently studied case of a disk with a fixed total mass, in which the viscosity is the primary variable that controls the disk temperature (*e.g.*, *Lunine and Stevenson*, 1982). Eqs. (4.5) and (4.6) provide an important constraint on the conditions needed to produce the Galilean satellites, because they define a maximum inflow rate consistent with temperatures low enough for forming ice-rich satellites.

Figure 5a shows the mid-plane temperature as a function of the assumed inflow rate from solution of equations (4.1), (4.3), and (4.4). Disk temperatures low enough for water ice (which is stable below ~ 150 to 200 K, depending on pressure, *e.g.*, *Prinn and Fegley*, 1989; see also *Canup and Ward*, 2002) in the $r \geq 15 R_J$ region require $\mathscr{F} \leq 6 \times 10^{-6}$ $M_\oplus$ yr$^{-1}$ for the parameter choices in Fig. 5a. A solar composition inflow having this inflow rate would take $\geq 10^6$ yr to supply a mass in solids comparable to that of the Galilean satellites. Fig. 5a shows that temperatures remain too high for ices in the innermost circumjovian disk even for very slow inflow rates, consistent with Io's anhydrous composition. While $T$ does not vary strongly with the choice of disk viscosity, the resulting gas surface density varies as $1/\nu$, so that an inflow with $\mathscr{F} = 10^{-5}$ $M_\oplus$ yr$^{-1}$ would for $10^{-2} \leq \alpha \leq 10^{-4}$ produce a disk having $\sigma_g \sim 10$ to $10^3$ g/cm$^2$ in the satellite region (Fig. 5b). A dusty disk having these surface densities would be optically thick. The corresponding disk pressures for the Fig. 5b curves range from ~ $10^{-5}$ to $10^{-2}$ bars in the $r < 30 R_P$ region. These are much lower surface densities and pressures than that of an MMSN disk, with $\sigma_{g, MMSN} \sim$ few $\times 10^5$ g/cm$^2$, and characteristic pressures ~ 1 to 10 bars (*e.g.*, *Prinn and Fegley*, 1989). A disk with a low, gaseous opacity and a low viscosity ($\alpha = 10^{-4}$) has ice stability for $r \geq 15 R_J$ if $\mathscr{F} < 5 \times 10^{-5}$ $M_\oplus$ yr$^{-1}$, implying that $> 10^5$ yr would be needed to supply a mass in solids equal to that of the Galilean satellites, assuming a solar composition inflow.

Because the satellite growth rate is regulated by the rate that solids are supplied to the disk (see section 4.2.3), a maximum inflow rate consistent with icy Ganymede and Callisto corresponds to a minimum satellite accretion timescale. Across a wide range of disk parameters, satellite formation timescales of order $10^5$ to $10^6$ yr are then implied. Such timescales seem reasonable if the final stages of inflow were regulated by the dispersal of the solar nebula.

### 4.1.3 *Discussion.*

In an actively supplied accretion disk, the inflow rate is the primary regulator of disk temperature. Temperatures in the region of the outer Galilean satellites would have become cold enough for ices only once the rate of inflow slowed substantially compared to its peak. This implies that the Galilean satellites 1) formed very late in Jupiter's gas accretion, 2) originated within an orders-of-magnitude lower gas density disk than suggested by MMSN models, and 3) accreted very slowly (in $> 10^5$ yr for a solar composition inflow ; see also § 4.2.3). While the third condition is potentially consistent with an undifferentiated Callisto (per discussion in § 2.4), we emphasize that in the *Canup and Ward* model, the requirement of slow accretion emerges primarily from the ice-rich compositions of Ganymede and Callisto, and is not directly dependent on Callisto's interior state.

As we discuss in § 4.3, it is certainly possible that earlier generations of satellites accreted (and were lost) during periods of faster gas inflow. Higher disk gas densities and temperatures during those periods suggest that such satellites would have been rock-rich, rather than icy (*Canup and Ward*, 2002; *Alibert et al.*, 2005).

## 4.2 Delivery and evolution of disk solids

### 4.2.1 *Sources of solids.*

To form satellites, rock and ice must be supplied to the accretion disk in addition to the gas. Possible sources of solid material include 1) direct transport of small particles into the disk with the inflowing gas, 2) capture of Sun-orbiting planetesimals as they pass through the disk and lose energy due to gas drag, and 3) capture of collisional debris from collisions between planetesimals within the planet's Hill sphere, or between a planetesimal and a planet-orbiting object.

Mechanism (1) applies for particles small enough to be aerodynamically coupled to the gas. The behavior of a Sun-orbiting particle of radius $r_p$ and density $\rho_p$ is controlled by the motion of the gas rather than by solar gravity if its so-called "stopping time" due to interaction with the gas, $t_e \equiv r_p \rho_p /(c \rho_g)$, is shorter than its orbital timescale, $\Omega_p^{-1} \equiv (GM_\odot / r^3)^{-1/2}$ (*e.g.*, *Weidenschilling and Cuzzi*, 1993). This is true for particles with radii

$$r_p << \frac{\rho_{g,n} c}{\rho_p \Omega_p} \approx \frac{\sigma_{g,n}}{2\rho_p} \sim O(1) \, \text{m} \left( \frac{\sigma_{g,n}}{400 \, \text{g/cm}^2} \right) \left( \frac{1.5 \, \text{g/cm}^3}{\rho_p} \right), \qquad (4.7)$$

where vertical hydrostatic equilibrium (so that $H \approx c/\Omega$ and $\sigma_{g,n} = 2H\rho_{g,n}$) has been assumed, and $\rho_{g,n}$ and $\sigma_{g,n}$ are the solar nebula density and surface density. Particles smaller than meter-sized would be delivered to the disk by the inflowing gas with a specific angular momentum similar to the gas itself, so that their initial orbits would have $r \leq r_o$. For $r_o \sim 30R_J$ (as estimated in § 3.1), this region would be comparable to the radial scale of the Galilean system.

To be captured via mechanisms (2) or (3), a particle on an initially heliocentric orbit must lose sufficient energy while in the presence of the planet to become planet-bound. One potential source of dissipation is gas drag due to passage through the circumplanetary disk. To be captured during a single passage through the disk, a particle must encounter a mass in gas comparable to or greater than its own; this would occur for $4\pi r_p^2 \sigma_g \geq 4\pi r_p^3 \rho_p / 3$, or for particle radii

$$r_p \leq 2 \text{ m } \left( \frac{\sigma_g}{10^2 \text{ g/cm}^2} \right) \left( \frac{1.5 \text{ g/cm}^3}{\rho_p} \right) . \qquad (4.8)$$

Given expected values for $\sigma_g$ (Fig. 5b), particles small enough to be captured would in most cases also be small enough to be coupled to the nebular gas, and so would instead be delivered to the disk with the inflowing gas via mechanism (1). Gas-drag capture would supply solids across the full radial width of the accretion disk (*i.e.*, for $r < r_d$), and so if this had been a dominant source of solid material for regular satellite accretion it is difficult to explain the compactness of the Galilean system.

An alternative mode of capture that is independent of the presence of gas involves mutual collisions, either between heliocentric orbiting objects colliding within the planet's Hill sphere ("free-free" collisions), or between a free object and a pre-existing satellite (a "free-bound" collision). For discussion of both processes, the reader is referred to *Safronov et al.*, (1986), *Estrada and Mosqueira*, (2006), and the *Estrada et al.* chapter. In the context of the accretion disk model we address here, mass delivery from free-free collisions, which is quite inefficient, is likely to be less than that due to small particles delivered with the inflowing gas. *Estrada and Mosqueira* (2006) estimate a free-free delivery rate of $\sim 2 \times 10^{18}$ g/yr for a minimum mass nebula whose solids are contained in 1-m to 10-km objects, while direct inflow of meter-sized and smaller particles with the gas can provide $\sim 6 \times 10^{20}$ g/yr $\left( \mathscr{F} / 10^{-5} M_\oplus \text{yr}^{-1} \right) \left( 10^2 / f \right)$.

However free-bound collisions could provide an additional source of solid material to the $r < r_o$ region once satellites have begun to form there.

Although both direct inflow and capture processes would have been ongoing as Jupiter's satellites formed, direct inflow of small particles with the gas appears the mode of solid delivery most naturally able to account for the radial scale of the satellite system. In general, all of the mechanisms that deliver solids to circumplanetary orbit become more effective as the assumed number density of small objects in the solar nebula is increased. Small bodies may be continually generated in late stage planet formation, because the growing planets stir neighboring planetesimals to high velocities so that collisions between planetesimals become fragmentary (*e.g.*, *Goldreich et al.*, 2004; *Rafikov*, 2004; *Chiang et al.* 2007; *Schlecting and Sari*, 2007).

### 4.2.2 *Retention of small particles.*

The initial evolution of small disk particles will be governed by a balance between accretional growth and orbital decay due to aerodynamic gas drag. The rate of gas drag orbital decay is inversely proportional to particle radius for particles with $t_e \geq \Omega^{-1}$ (*i.e.*, for sizes comparable to and larger than the particle size most strongly coupled to the gas), so that if particles grow larger quickly enough, they can avoid decay. The decay timescale, $\tau_{gd} \equiv r/\dot{r}$, for a radius $r_p$ and density $\rho_p$ particle at orbital distance $r$ is $\tau_{gd} \sim (8/3C_D)(\rho_p r_p / \sigma_g)(r\Omega/c)^3 \Omega^{-1}$, where $C_D$ is the drag constant of order unity for particles too small to gravitationally perturb the gas. The timescale for a mass $m_p$ and radius $r_p$ particle to grow by accretion, $\tau_{acc} \equiv m_p / \dot{m}_p$, (or alternatively, the time spent as a mass $m_p$ particle), is $\tau_{acc} \sim \rho_p r_p /(\sigma_p \Omega F_G)$. Here $\sigma_p$ is the surface density of disk particles, and $F_G \equiv (1 + v_{esc}^2 / v^2)$ is the gravitational focusing factor, a function of the ratio between the mutual escape velocity of colliding particles, $v_{esc}$, and their relative velocities, $v$.

Comparing $\tau_{acc}$ to $\tau_{gd}$ gives (*e.g., Makalkin et al.* 1999; *Canup and Ward*, 2002)

$$\frac{\tau_{acc}}{\tau_{gd}} \sim 10^{-3} \left( \frac{5}{F_G} \right) \left( \frac{\sigma_g / \sigma_p}{100} \right) \left( \frac{c/r\Omega}{0.1} \right)^3 , \qquad (4.9)$$

independent of particle size. When this ratio is less than unity, particles can grow through mutual collisions faster than they are lost due to gas drag. The term in Equation (4.9) that is least certain is the disk gas-to-solid surface density ratio. For a solar composition mixture, $(\sigma_g/\sigma_p) \sim$ $10^2$, and even a disk that was one to two orders-of-magnitude more gas-rich would still have $(\tau_{acc}/\tau_{gd}) \ll 1$. When $(\tau_{acc}/\tau_{gd}) \ll 1$, solids are expected to accrete across the $r < r_o$ region where they are delivered by the inflow.

While the inflowing gas viscously spreads and maintains a quasi-steady state surface density, the inflowing solids will effectively be decoupled from the gas once they begin to accrete. This means that solids will build-up in the circumplanetary disk and that the gas-to-solids ratio in the disk will be lower than in the inflowing material. For example, spreading the mass of the Galilean satellites into a disk extending to $30R_P$ gives an average solid surface density $\sigma_p \sim 3 \times 10^3$ g/cm$^2$. Requiring that the inflow rate was slow enough for ices in the Callisto-Ganymede region implies a disk gas surface density $< 10^3$ g/cm$^2$ for $\alpha > 10^{-4}$ (Fig. 5b), implying a circumplanetary disk gas-to-solids ratio $(\sigma_g/\sigma_p) \leq 1$, and that the Galilean satellites formed in gas-starved conditions.

### 4.2.3  *Rate of satellite growth.*

The satellite accretion rate is a control on the initial thermal states of the satellites and is key to explaining an incompletely differentiated Callisto. In an MMSN disk, all of the material needed to form the satellites is presumed to be in the disk as a starting condition. In this case, the time needed to accrete an $R_S \sim 2500$ km, $\rho_S \sim 1.8$ g/cm$^3$ Callisto-like satellite is $\tau_{acc} \sim \rho_S R_S /(\sigma_p \Omega F_G) \sim 600$ yr $(10^3$ g/cm$^2 / \sigma_p) (r / 26 R_J)^{3/2} (5 / F_G)$. As discussed in § 2.4, with such rapid accretion it appears impossible to avoid melting Callisto's ice, because peak temperatures in the satellite—even in the limit of maximally effective cooling—reach $\sim 10^3$ K (*e.g.*, *Stevenson et al.*, 1986).

However, in an accretion disk the formation time of large satellites can be much longer, because it is regulated by the supply rate of particles to the disk. For a gas inflow rate slow enough to allow ices in the Ganymede-Callisto region (*i.e.*, $\mathscr{F} \sim 10^{-5}$ $M_\oplus$ yr$^{-1}$ from § 4.1.2), the time required for the inflow to deliver a mass in solids equal to that of the Galilean satellites, $M_T = 2.1 \times 10^{-4}$ $M_J$, is $\tau_{in} \approx (f / \mathscr{F}) M_T \sim 7 \times 10^5$ yr $(f / 10^2)(10^{-5} M_\oplus$ yr$^{-1} / \mathscr{F})$, where $f$ is the gas-to-solids ratio in the inflowing material. Such long formation times are comparable to those required to explain Callisto ($\tau_{acc} > 5 \times 10^5$ yr; *Barr and Canup*, 2008). In general, in an inflow-sustained disk slow satellite growth arises directly from the same conditions necessary to produce ice-rich satellites, and accounting for Callisto's apparent interior state does not require the imposition of additional constraints.

## 4.3 Growth and loss of large satellites

Just 2% of Jupiter's mass in solar composition material is needed to produce the Galilean satellites, and yet a much greater total mass – perhaps as much as 10% or more of Jupiter's mass – may well have been processed through the Jovian accretion disk (*e.g.*, § 3). The latter would imply a total mass in inflowing solid material that is many times larger than that contained in the Galilean satellites. Once inflowing particles begin to accrete, they are likely to grow faster than they would be lost to aerodynamic gas drag (equation (4.9)), implying a high efficiency for their incorporation into satellites. Why then aren't the Galilean satellites Mercury or even Mars-sized? *Canup and Ward* (2006) proposed that Type I migration precludes the formation of satellites larger than a critical mass, and that migration will effectively select for the formation of satellite systems around gas planets containing a total mass ~ $O(10^{-4})$ times the planet's mass, as we now describe.

### 4.3.1 *Satellite orbital migration: Type I and Type II.*

The effects of disk torques that arise from the interaction of an orbiting object with a companion gas or particle disk have received increasing attention in recent decades. The discovery of "hot Jupiters" in extrasolar systems –*i.e.*, Jupiter-like planets that occupy orbits extremely close to their central stars – seems to require that gas giant planets were able to migrate inward over large distances. Disk torques provide a natural means for inward planet migration and even the loss of planets due to collision with the central star. Analogous satellite-disk torques have important implications for Galilean satellite formation.

As a satellite grows, its gravitational interactions with a background gas disk become increasingly important, and these interactions lead to torques on the satellite. The satellite's gravity induces spiral density waves in the gas disk, which are initiated at Lindblad resonant locations both interior and exterior to the satellite and propagate as pressure waves (*e.g.*, *Goldreich and Tremaine*, 1980). The torque on the satellite due to the exterior waves leads to a decrease in the satellite's orbital angular momentum, while the interior waves lead to a positive torque on the satellite. Because the total torque due to the exterior waves is of order 10% larger than that of the interior waves, the net torque on the satellite is negative and acts as a drag on the satellite's orbit, causing it to slowly spiral inward towards the planet through a process known as Type I migration (*e.g.*, *Ward*, 1986; *Ward*, 1997).  The density wave torque, $\dot{L}_S$, is proportional to $M_S{}^2$, while the satellite's orbital angular momentum, $L_S$, is proportional to $M_S$, so that the timescale over which $L_S$ changes is $L_S / \dot{L}_S \propto 1/M_S$, and migration becomes more rapid as a satellite grows.

The Type I migration timescale is

$$
\begin{aligned}
\tau_1 &\equiv \frac{r}{\dot{r}} = \frac{1}{C_a \Omega} \left( \frac{M_P}{M_S} \right) \left( \frac{M_P}{r^2 \sigma_g} \right) \left( \frac{c}{r\Omega} \right)^2 \\
&\sim 4 \times 10^5 \ \mathrm{yr} \left( \frac{3.5}{C_a} \right) \left( \frac{50 \ \mathrm{g/cm^2}}{\sigma_g} \right) \left( \frac{10^{26} \ \mathrm{g}}{M_S} \right) \left( \frac{30}{r/R_J} \right)^{1/2} \left( \frac{c/r\Omega}{0.1} \right)^2
\end{aligned}
\tag{4.10}
$$

where $M_P$ is the planet's mass, $\Omega \equiv (GM_P / r^3)^{1/2}$, $(c/r\Omega) \sim 0.1$ and $C_a$ is a constant (*Tanaka et al.*, 2002).  Approximate values appropriate for Galilean satellite accretion within a slow-inflow produced disk (*e.g.*, Figure 5b) are given in the second expression.  Density wave interactions also circularize satellite orbits, with an eccentricity (*e*) damping timescale, $\tau_e = e/\dot{e} \sim \tau_1 (c/r\Omega)^2$, which is much shorter than the Type I timescale (*e.g.*, *Ward*, 1988; *Artymowicz*, 1993; *Papaloizou and Larwood*, 2000; *Tanaka and Ward*, 2004).

For a sufficiently massive satellite, the response of the disk to its density wave perturbations becomes non-linear, and the satellite clears an annular region of the disk centered at its orbital radius, producing a "gap" of reduced surface density compared to that of the background disk.  A satellite massive enough to open a gap will be subject to Type II migration, in which its orbital evolution is coupled to the viscous evolution of the disk (*e.g.*, *Lin and Papaloizou*, 1986).  A standard criterion for gap opening and transition to Type II migration is that the cumulative torque on the disk by the satellite must exceed the rate of angular momentum transport due to the disk's viscosity, so that the satellite repels local disk material faster than it can be re-supplied by viscous diffusion (*e.g.*, *Lin and Papaloizou*, 1986; *Ward*, 1986).  If the waves are damped locally, a gap-opening criterion that considers an alpha viscosity model is (*e.g.*, *Ward and Hahn*, 2000)

$$
\frac{M_S}{M_P} > c_v \sqrt{\alpha} \left( \frac{c}{r\Omega} \right)^{5/2} \sim 2 \times 10^{-4} c_v \left( \frac{\alpha}{3 \times 10^{-3}} \right)^{1/2} \left( \frac{c/r\Omega}{0.1} \right)^{5/2}
\tag{4.11}
$$

where $c_n$ is a constant of order unity to ten. For $c_n = 3$ and $(c/r\Omega) = 0.1$, the mass cutoff in equation (4.11) is larger than Ganymede's mass for $\alpha > 7 \times 10^{-5}$. If a satellite grows large enough to open a gap, Type II migration will generally also lead to orbital decay with a timescale comparable to the local disk's viscous spreading time,

$$\tau_{II} \approx \tau_\nu = \frac{r^2}{\nu} \approx 300 \text{ yr} \left(\frac{3\times10^{-3}}{\alpha}\right)\left(\frac{r}{30R_J}\right)^{3/2}\left(\frac{0.1}{c/r\Omega}\right)^2. \qquad (4.12)$$

### 4.3.2 *Maximum satellite mass.*

A satellite will grow through the sweep-up of inflowing solids until it reaches a critical mass for which the characteristic time for its further growth is comparable to its Type I orbital decay timescale. Satellites cannot grow substantially larger than this critical mass before they are lost to collision with the planet, because Type I migration becomes more rapid as a satellite's mass increases.

Consider a satellite at orbital distance $r$ that accretes material across an annulus of width $\Delta r$ in an inflow-supplied accretion disk. The satellite's growth timescale is $\tau_{acc} \sim fM_S / (2\pi r \Delta r f_{inflow})$, where $f_{inflow}$ is the total mass inflow rate per area (gas and solids), and $f$ is the mass ratio of gas-to-solids in the inflow. The annulus width $\Delta r$ is set by the larger of two quantities: the distance over which a satellite can gravitationally perturb particles into a crossing orbit with itself (which is proportional to the satellite's Hill radius) or the radial excursion executed by an eccentric satellite during a single orbit, with $\Delta r \sim 2er$. From numerical simulations (see § 4.4), *Canup and Ward* (2006) found that $\Delta r$ is set by the characteristic maximum eccentricity, $e$, that results from a balance between eccentricity damping by density waves and excitation via gravitational scatterings with similarly sized objects (*e.g.*, *Ward*, 1993), with $e \sim (c/r\Omega)(M_S/4\pi rH\sigma_g)^{1/5}$. Using this estimate together with $\Delta r \sim 2er$ gives the satellite accretion timescale, $\tau_{acc} \sim (f\sigma_g / f_{inflow})(M_S / 4\pi rH\sigma_g)^{4/5}$.

The critical maximum satellite mass, $M_{crit}$, is then found by setting $\tau_{acc} \sim \tau_1$ from equation (4.10), using $\sigma_g$ from equation (4.1), and solving for $M_S = M_{crit}$. Defining $\tau_G \equiv M_P / \mathscr{F}$ as the time in which a mass equal to the planet's mass is delivered by the inflow, the critical satellite mass in planet masses is (*Canup and Ward*, 2006)

$$\left(\frac{M_{crit}}{M_P}\right) \approx 5.4 \left(\frac{\pi}{C_a}\right)^{5/9} \left(\frac{c}{r\Omega}\right)^{26/9} \left(\frac{r}{r_o}\right)^{10/9} \left(\frac{\alpha}{f}\right)^{2/3} \left[\Omega\tau_G f\right]^{1/9}$$

$$\sim 5.6\times10^{-5} \; \chi \; \left(\frac{3.5}{C_a}\right)^{5/9} \left(\frac{c/r\Omega}{0.1}\right)^{26/9} \left(\frac{r/r_o}{0.5}\right)^{10/9} \left(\frac{\alpha/f}{3\times10^{-5}}\right)^{2/3} \tag{4.13}$$

where $\chi \equiv \left[(1\,\text{week}/\{2\pi/\Omega\})(f/10^2)(\tau_G/10^7\,\text{yr})\right]^{1/9}$ is a factor close to unity. The Galilean satellites have $(M_S/M_P)$ values ranging from $2.5 \times 10^{-5}$ to $7.8 \times 10^{-5}$, comparable to the critical mass in equation (4.13). The predicted $(M_{crit}/M_P)$ ratio depends very weakly on the inflow rate through the $(\tau_G)^{1/9}$ term in $\chi$, which implies that a similar maximum satellite mass would be expected across a wide range of inflow rates. The quantity $(c/r\Omega)$ varies weakly with $r$ for most disks, *e.g.,* varying from $\sim 0.09$ to $0.11$ across the $R_P < r < 30R_P$ region for the inflow conditions shown in Fig. 5b. If satellites form throughout the inflow region, the ratio $(r/r_o)$ will be of order 0.1 to unity. The last term in equation (4.13) contains the ratio of two uncertain parameters: the $\alpha$ viscosity parameter and $f$. For a given inflow rate, a higher viscosity yields lower disk gas surface densities, and thus allows larger mass satellites to survive against Type I decay, while a lower $f$ implies a more solid-rich inflow, which speeds up satellite growth allowing objects to grow larger before they are lost.

The above assumes that satellites undergo Type I migration and do not grow large enough to open gaps. Comparing the gap opening mass, $M_{gap}$, from equation (4.11) to $M_{crit}$ from (4.13) gives

$$\frac{M_{gap}}{M_{crit}} \sim 9 \left(\frac{c_\nu}{3}\right) \left(\frac{C_a}{3.5}\right)^{5/9} \left(\frac{3\times10^{-3}}{\alpha}\right)^{1/6} \left(\frac{0.1}{c/r\Omega}\right)^{7/18} \left(\frac{0.5}{r/r_o}\right)^{10/9} \left(\frac{f}{10^2}\right)^{2/3}. \tag{4.14}$$

Unless the disk is extremely viscous and/or the inflow is substantially more solid-rich than solar composition, $M_{gap}/M_{crit} > 1$, and satellites will be lost to Type I decay before they grow large enough to open gaps and transition to Type II migration. In the satellite systems produced by the *Canup and Ward* (2006) accretion simulations (§ 4.4), the final satellites were all less massive than $M_{gap}$, with the largest satellite in each system having an average mass $\sim 0.2 \pm 0.1 M_{gap}$ for $c_n$ = 3.

### 4.3.3  *Creation of common $(M_T/M_P)$.*

The limiting satellite mass in (4.13) also implies a limit on the total mass of a satellite system. Consider an inflow that persists for a time exceeding that needed for satellites of mass $M_{crit}$ to form. Within a given annulus in the disk, a satellite grows to a mass $\sim M_{crit}$ before being lost to Type I decay, but in a comparable timescale to its demise another similarly massive satellite grows in its place, because $\tau_1 \sim \tau_{acc}$. In this way, the disk is regulated to contain a total mass in satellites, $M_T$, comparable to a distribution of mass $M_{crit}$ objects across the inflow region. For ($c/r\Omega$) and $f$ that are approximately constant across the disk, the predicted satellite system mass fraction, ($M_T/M_P$), is (*Canup and Ward*, 2006)

$$\left(\frac{M_T}{M_P}\right) = \int_{r_i \sim R_p}^{r_o} \frac{(m_{crit}/M_p)}{\Delta r} dr \approx 2.1 \left(\frac{\pi}{C_a}\right)^{4/9} \left(\frac{c}{r\Omega}\right)^{10/9} \left(\frac{\alpha}{f}\right)^{1/3} \frac{1}{[\Omega\tau_G f]^{1/9}}$$

$$\sim 2.5\times10^{-4} \ \frac{1}{\chi}\left(\frac{3.5}{C_a}\right)^{4/9} \left(\frac{c/r\Omega}{0.1}\right)^{10/9} \left(\frac{\alpha/f}{3\times10^{-5}}\right)^{1/3}, \quad (4.15)$$

similar to ($M_T/M_P$) for both the Jovian and Saturnian systems. The fraction ($M_T/M_p$) is insensitive to inflow rate through $\chi$, lacks any dependence on $r_o$, and depends quite weakly on ($\alpha/f$) to the one-third power. The implication is that satellite systems formed within accretion disks of gas planets will have ($M_T/M_P$) $\sim O(10^{-4})$ across a wide range of inflow and disk parameters, so long as the disk is viscous and Type I migration is active. This offers an explanation for why the satellite systems of Jupiter and Saturn would contain a similar fraction of their central planet's mass, even though the individual accretion histories of these two planets (*e.g.*, gas inflow rates, total mass processed through their accretion disks, etc.) may well have differed.

### 4.3.4 Extent of migration of Galilean satellites

Initially a satellite's mass accretion timescale is short compared to its Type I migration timescale. However once its mass exceeds $M_{crit}$ given in equation (4.13), these timescales invert. For a disk produced by a uniform inflow per area having $\sigma_g \propto 1/r^{3/4}$ and ($c/r\Omega$) $\propto r^{1/8}$ (*Canup and Ward*, 2002), equation (4.10) gives $\tau_I \propto r^{3/2}/M_S$. Once a satellite mass exceeds $M_{crit}$ it migrates inward faster than it grows, and its characteristic migration timescale shortens as $r$ decreases and $M_S$ increases.

It seems probable that earlier generations of satellites formed and were then lost to collision with Jupiter by this process. The final Galilean satellites must have avoided this runaway migration phase, implying that their final masses were comparable to or smaller than $M_{crit}$ in their precursor disk at the time the inflow ended and the disk dissipated. However they would still have migrated inward somewhat as they formed. The satellite accretion rate (assuming a uniform inflow per area; *Canup and Ward*, 2006; also § 4.3.2) and the Type I migration rate are

$$\frac{dM_S}{dt} \sim M_S \left( \frac{f_{inflow}}{f \sigma_g} \right) \left( \frac{4\pi r H \sigma_g}{M_S} \right)^{4/5}$$
$$\frac{dr}{dt} \sim -r\Omega C_a \left( \frac{M_S}{M_P} \right) \left( \frac{r^2 \sigma_g}{M_P} \right) \left( \frac{r\Omega}{c} \right)^2 . \qquad (4.16)$$

Together these give the change in satellite mass as a function of radius, $dM_S / dr = (dM_S / dt)/(dr / dt)$, which can be directly integrated under the assumption that the gas surface density and the normalized disk scale height can be approximated as $\sigma_g = \sigma_{g,0} (R_P / r)^{\gamma_g}$ and $(c/r\Omega) = (c/r\Omega)_0 (r/R_P)^{\gamma_c}$, where the terms subscripted zero refer to values at $r = R_P$. For $f_{inflow} \equiv \mathscr{F} /(\pi r_0^2)$, and $\sigma_{g,0}$ and $(c/r\Omega)_0$ determined from equation (4.1), the initial satellite orbital radius $r_0$ as a function of the final satellite mass ($M_S$) and final position ($r_f$) is

$$\frac{r_0}{R_P} = \left[ K(\beta+1)\left( \frac{M_S}{M_P} \right)^{9/5} + \left( \frac{r_f}{R_P} \right)^{\beta+1} \right]^{1/(\beta+1)} , \qquad (4.17)$$

where $\beta \equiv (1/10) + (6/5)\gamma_g + (14/5)\gamma_c$, and

$$K \cong 2.16 \times 10^9 \left( \frac{C_a}{3.5} \right) \left( \frac{1.25 - \sqrt{r_o / r_d}}{0.8} \right)^{6/5} \left( \frac{r_0 / R_P}{30} \right)^2 \left( \frac{3 \times 10^{-5}}{\alpha / f} \right) \left( \frac{3 \times 10^{-3}}{\alpha} \right)^{1/5} \left( \frac{\mathscr{F}}{10^{-5} \, M_\oplus \text{yr}^{-1}} \right)^{1/5} \left( \frac{0.1}{(c/r\Omega)_0} \right)^{26/5}$$
$$(4.18)$$

For a uniform inflow per area, *Canup and Ward* (2002) find $\gamma_g \sim 3/4$ and $\gamma_c \sim 1/8$ in the inner disk, or $\beta = 27/20$. With this $\beta$, the nominal values given in (4.18), and the current satellite masses and orbital radii, the predicted starting radii for Io, Europa, Ganymede and Callisto from (4.17) are then $\sim 8$, 10, 17 and 27$R_J$, respectively.

Disk temperatures exterior to ~ $20R_J$ drop below the ice stability line once the inflow rate has declined to about $\mathscr{F} < 10^{-5}$ $M_\oplus$ yr$^{-1}$ for the parameter choices in Fig. 5a, consistent with Ganymede and Callisto's ice rich compositions including effects of inward migration during their formation. The simple migration estimate in equation (4.17) neglects interactions between the growing satellites, including scattering, resonances, and/or mutual collisions and mergers. Accretion simulations including such effects find that the outer large satellites migrate inward by up to about 5 planetary radii in those systems that most closely resemble the Galilean satellites (*Canup and Ward*, 2006, e.g., their Fig. 3b).

## 4.4  Direct accretion simulations

Relatively few works have modeled Galilean satellite accretion. *Coradini et al.* (1989) analytically estimated satellite masses and orbital separations assuming a variety of values for the gravitational focusing factor. *Richardson et al.* (2000) used direct *N*-body simulations to model accretion within an MMSN disk, tracking the growth of satellites containing up to ~ 10% of the mass of Io. *Canup and Ward* (2002) and *Alibert et al.* (2005) considered satellite growth and loss due to Type I migration within an accretion disk by assuming a linear growth in satellite mass with time.

*Canup and Ward* (2006) presented the first direct simulations of the accretion of Galilean-sized satellites. They used an *N*-body integration (*Duncan et al.*, 1998) modified to include the effects of ongoing inflow to the disk and interactions between the growing satellites and the gas disk. To mimic the inflow of solids, orbiting objects were added to the disk with random position within $r < r_o$ at a rate proportional to $(\mathscr{F}/f)$. Computational limitations on the number of objects dictated that the added objects were large, typically ~ $10^2$ km. All objects were tracked with direct orbit integration, and collisions were assumed to result in the inelastic merger.

Figure 6 shows results of three accretion simulations that assumed a constant rate of gas inflow and different values for ($\alpha/f$). Initially the inflow of solids causes the total mass in orbiting satellites, $M_T$, to increase with time until objects of mass ~ $M_{crit}$ form (equation (4.13)). The orbits of the largest satellites then begin to decay inward due to Type I migration, and $M_T$ decreases in discrete steps as satellites are removed due to collision with Jupiter. Continuing solid inflow to the disk leads to the growth of another generation of mass ~ $M_{crit}$ satellites on a timescale comparable to that of the prior satellites' orbital decay. The net result is a repeating cycle of satellite formation and loss. At any time beyond that needed to initially form the first mass $M_{crit}$ satellites, the total mass contained in satellites relative to that in the planet, $(M_T/M_P)$, oscillates about a nearly constant value similar to that in equation (4.15). The analytical estimates in (4.13) and (4.15) consider disk annuli independently. But in the simulations, as satellites formed in the outer disk migrate inwards they pass through interior zones and accrete additional material along the way. Migration-driven growth hastens their orbital decay, so that they are lost somewhat more quickly than the time to replenish their mass in their original radial zone. This causes the $(M_T/M_p)$ value from (4.15) to be an approximate upper limit, as seen in Fig. 6. Because this predicted satellite system mass fraction is proportional to $(\alpha/f)^{1/3}$, the factor of 500 variation in $(\alpha/f)$ considered across the Fig. 6 simulations yields only a factor of $\sim 10$ spread in $(M_T/M_p)$.

The actual inflow to Jupiter would have been time-dependent. In a system with an exponentially decaying inflow, the final system of satellites arises when the satellite growth timescale becomes comparable to the inflow decay timescale. As the inflow wanes and the gas disk disperses, Type I migration ends and a final system of satellites stabilizes. Figure 7 shows results of simulations with time-dependent inflows that were assumed to decay exponentially with time constants between $2 \times 10^5$ and $2 \times 10^6$ yr, for a variety of inflow and disk parameters. Many cases produce systems whose number of large satellites and satellite masses are similar to those of the Galilean system. The median number of large, $(M_S/M_P) \geq 10^{-5}$ final satellites in the *Canup and Ward* (2006) simulations was four, and the simulated satellites had average orbital separations comparable to those of the Galilean satellites. The radial extent of the system is set predominantly by the choice of $r_o$, with the most distant large satellite typically having a semi-major axis $a_{max} \sim r_o$, with this value varying by about a factor of 2 for a fixed value of $r_o$ due to dynamical scattering and satellite migration (*e.g.,* Fig. 7a, triangles vs. diamonds).

*Canup and Ward* (2006) showed that similar accretion conditions can also yield systems in which a single satellite contains most of the final satellite system mass, such as is the case for Titan and the Saturnian satellites. This outcome occurs in an initially Galilean-like system when the inner satellites suffer a shorter characteristic migration timescale than the outermost satellite. If the gas inflow to the disk ends after the inner large satellites have been lost to collision with the planet but before a new complete generation of satellites is formed, a single large satellite can be the sole survivor. For the Saturnian system, such a process would imply that the small and medium sized Saturnian satellites interior to Titan formed after all inner massive companions to Titan had been lost. Inflow rates during this final era could have been low enough to allow ice even in the inner Saturnian disk, consistent with the ice-rich compositions of Saturn's inner satellites (*Canup and Ward*, 2006; *Barr and Canup*, 2008).

In summary, the balance between inflow-regulated satellite growth and gas-driven satellite loss causes the ratio between the total satellite system mass and the planet's mass to maintain a roughly constant value, which is nearly independent of both the inflow rate and its characteristic angular momentum (which sets $r_o$). This common fractional value is weakly dependent on the ratio of the disk viscosity alpha parameter and the inflow's gas-to-solid composition (Figs. 6 and 7c), with $(M_T/M_P) \propto (\alpha/f)^{1/3}$, which implies that satellite system masses similar to those of Jupiter and Saturn result for $10^{-6} < (\alpha/f) < 5 \times 10^{-4}$, *i.e.*, a span of nearly 3 orders of magnitude. While quantitative predictions for $\alpha$ and $f$ are lacking, $(\alpha/f)$ ratios throughout this range are consistent with commonly assumed values for $\alpha$ and $f$, *e.g.*, $10^{-4} < \alpha < 0.1$ and $10 < f < 10^2$.

## 5    DISCUSSION

The accretion disk model reviewed here has a number of strengths. It describes satellite formation as a natural consequence of Jupiter's formation. Provided the planet contracted before nebular dissipation was complete, a circumplanetary accretion disk is expected, based on analytical estimates of the expected angular momentum of inflowing gas and on results of hydrodynamic simulations. If the disk is viscous, the outward angular momentum transport required to allow gas to be accreted by the planet is achieved. Between the period of peak runaway gas inflow expected for a gas giant planet and the cessation of gas accretion, inflow rates to the circumjovian disk would slow, allowing for disk temperatures low enough for ice stability in the general region of Ganymede and Callisto, with ice accretion at Europa's location possible late in the satellite's growth. The slow inflow rates needed for ice-rich satellites in turn imply that the Galilean satellites would have accreted slowly, in keeping with conditions needed for a Callisto-sized satellite to avoid melting and differentiation as it forms. Estimates of the average specific angular momentum of inflowing gas are broadly consistent with the radial scale of the Galilean system. Direct accretion simulations show the formation of Galilean-like satellite systems for a range of disk and inflow parameters, and suggest that Type I migration would regulate gas planet satellite systems to contain a common fraction of the planet's mass, independent of the total mass processed through the protosatellite disk. This offers an attractive explanation for the similarity of the Galilean and Saturnian satellite system masses compared to their planet's mass.

However, some aspects of the accretion disk model are not well understood. Foremost is the origin of viscosity in the protosatellite disk, and indeed the origin of viscosity in the cool and/or dusty regions of the surrounding circumsolar disk as well. It is generally accepted that magnetorotational instability (MRI) in weakly magnetized disks drives turbulence and outward angular momentum transport, with effective $\alpha$-values $\sim 10^{-2}$ to 0.1 (*Balbus and Hawley*, 1991). To be operative, MRI requires a minimum ionization fraction, which can be produced thermally (for temperatures in excess of $\sim 10^3$ K), by galactic cosmic-rays (for regions of low gas surface density with $\sigma_g < 10^2 \, \text{g/cm}^2$, *e.g.*, *Dullemond et al.* 2007), or by stellar X-rays (*e.g., Glassgold et al.*, 1997). However, the presence of dust grains can deactivate MRI because they are effective charge absorbers and can cause the ionization fraction to plummet (*Sano et al.* 2000). It is thus plausible that both the protosatellite disk and the circumsolar disk at Jupiter's distance would have been "layered": comprised of MRI "active" regions at the disk surfaces, together with a neutral, dust-rich "dead zone" surrounding the disk mid-plane (*e.g.*, *Gammie*, 1996). It is challenging to determine the exact vertical and radial extent of the inner dead zone, because this depends on the disk chemistry and the dust grain size distribution (*e.g.*, *Semenov et al.*, 2004), however it is possible that in a gas-starved protosatellite disk the active region could be substantial, and this merits investigation.

Non-magnetic hydrodynamic mechanisms have also been proposed as sources of turbulence and viscous transport in neutral disks, including notably those that rely on non-axisymmetric effects, including decaying vortices (*e.g.*, *Lithwick*, 2007), baroclinic instability (*Klahr and Bodenheimer*, 2003), and shear instability (*Dubrulle et al.* 2005). Recent experimental results claim to rule out angular momentum transport via axisymmetric hydrodynamical turbulence at levels corresponding to $\alpha > 10^{-5}$ (*Hantao et al.*, 2006). Gravitational torques from objects accreting within the disk may provide an effective viscosity (*Goodman and Rafikov*, 2001; see also *Canup and Ward*, 2002). In the specific context of an actively supplied accretion disk, the shock front resulting from the difference in the free-fall velocity of mass infalling to the disk and that of the orbiting gas is another potential source of turbulence and therefore viscosity (*Cassen and Moosman*, 1981). Such an inflow-driven viscosity would cease once the inflow had stopped completely, but it is possible that the last remaining orbiting gas could be then removed either by other sources of viscosity, *e.g.* by MRI once small grains have been removed through accretion onto the satellites. Spiral shocks generated by the inflowing gas may also drive angular momentum transport (*Lubow et al.*, 1999), although such shocks appear weaker in 3D simulations than in 2D simulations (*Bate et al.*, 2003).

Another area of uncertainty is the nature of the inflowing material. Hydrodynamical simulations clearly show the formation of a viscous circumplanetary disk (*e.g.*, Fig. 1 from *Bate et al.*, 2003). A key open issue is the average specific angular momentum of the inflowing material (which controls $r_o$), as well as how the inflow varies with distance from the planet. *D'Angelo et al.* (2003) find a peak specific angular momentum of circumplanetary material about 2.6 times larger than that of material orbiting Jupiter at Callisto's distance. In contrast, simulations by *Machida et al.* (2008) and *Machida* (2008) that resolve Jupiter's Hill sphere with much higher resolution find an average inflow specific angular momentum comparable to that at Callisto's orbit, or $r_o \sim 22 R_J$. *Machida* (2008) argues that this result is unaffected by the more local nature of the *Machida et al.* simulations. Further work is needed to address this important issue. The mass fraction of gas vs. solid particles in the inflow (the "*f*" parameter) is highly uncertain, and is likely to remain so. The predicted fraction of nebular solids contained in small particles as a function of time is inherently model-dependent, as is the rate of nebular dispersal and therefore the nebular gas surface density as a function of time. This means that both the numerator and denominator needed to estimate the ratio $f$ are poorly constrained, with a wide range of values possible.

It has been suggested that a region at the outer edge of Jupiter's gap—within which there would be a positive gradient in gas density with orbital radius—may frustrate the delivery of small particles to Jupiter. *Rice et al.* (2006) argue that outward gas drag of small particles across this region would prevent all but very small dust particles from reaching the planet. However, this implies that small particles will build-up at a preferred location at the gap's outer edge. Once the local particle density exceeds that of the gas, the particle drift rate will drop rapidly (*e.g.*, *Nakagawa et al.*, 1986), making it more difficult to prevent particle diffusion into the planet's gap. This could result in a quasi-steady state flow interior to the region of particle build-up that has a similar solids content as that of the global disk flow, alleviating the problem. Assessing whether this occurs will require models that incorporate the effects of particle concentration. It is also possible that strong turbulence concentrated at the gap edges (*e.g.*, Ward, in preparation) will weaken such efficient particle filtering by gas-drag.

The accretion simulations of *Canup and Ward* (2006) found that overall outcomes were rather weakly dependent of the ratio of ($\alpha/f$), a fortuitous result given the uncertainties in both of these quantities discussed above. However, the smallest sized particles in the *Canup and Ward* simulations were $\sim 10^2$ km-scale due to computational limitations, so that the regime of small particle growth was not resolved. Future work should address this regime, and in particular the conditions in which particle accretion can occur faster than particle loss due to aerodynamic gas drag. The potential role of other solid delivery mechanisms in the context of an inflow-supplied accretion disk, including gas drag and collisional capture, should also be assessed.

## 6    IMPLICATIONS

In the past two decades there have been important advances in our understanding of satellite properties, gas giant planet formation, and planet and satellite migration processes. This chapter has focused on a model motivated in part by such advances, in which Europa and the other Galilean satellites form within a gas-starved accretion disk produced during the final stages of Jupiter's own gas accretion. Compared to the pioneering models developed in the 1980's, including most notably the work of *Lunine and Stevenson* (1982), the slow-inflow accretion disk construct implies notably different conditions for the disk environment in which Europa formed and for the nature of satellite accretion around gas planets. We here conclude by outlining overall implications of the *Canup and Ward* (2002, 2006) model.

1) *Satellite migration and loss.* If more than $\sim 2\%$ of Jupiter's mass in solar-composition material was processed through its accretion disk—which seems likely—multiple generations of satellites may have formed and ultimately been lost to Type I migration and collision with Jupiter. The Galilean satellites are then the last surviving generation, and their properties reflect conditions that existed as gas inflow to Jupiter was ending.

2) *Maximum satellite mass and a common total satellite system mass.* Type I migration precludes the formation of satellites larger than a critical mass, and causes the total satellite system mass to maintain a relatively constant value. While detailed characteristics of satellite systems of gas giants would vary, those that formed from accretion disks would be predicted to display a similar largest satellite mass and satellite system mass compared to their host planet's mass.

3) *Late formation times.* The growth of Europa and the other Galilean satellites is predicted to be concurrent with the dissipation of the solar nebula. This is likely to have occurred several million years or more after the formation of the earliest solar system solids, based on several lines of reasoning. Forming Jupiter via core accretion appears to require a few to ~ 5 million years (*Hubickyj et al.* 2005), and observations of disks around other stars find a mean nebular lifetime of 3 Myr (*Haisch et al.* 2001). The requirement that Callisto did not melt as it formed implies that the satellites finished their accretion no earlier than about 4 Myr after CAI's (*Barr and Canup*, 2008). These age estimates suggest that short-lived radioisotopes would have been relatively unimportant by the time the Galilean satellites accreted.

4) *Low gas disk densities, low disk pressures.* Accretion disks produced during the end stages of Jupiter's growth have low gas densities and pressures not too different from that of the background solar nebula. Infalling solid particles would therefore undergo little additional chemical processing, and their composition would be expected to generally reflect solar nebula conditions (*e.g.*, *Mousis and Alibert*, 2006). This contrasts to the dense and high-pressure disks implied by MMSN models, which implied substantial chemical alteration of disk material (*e.g.*, *Prinn and Fegley*, 1981).

5) *Protracted satellite accretion.* The slow inflow rates required for temperatures low enough for ices in the Ganymede and Callisto region imply that the Galilean satellites accreted slowly, taking $10^5$ to $10^6$ yr to acquire their final masses for a solar composition inflow. Such long accretion times are also generally consistent with those needed for Callisto to avoid melting as it accreted. In contrast, traditional MMSN models predict much more rapid satellite accretion in $\leq 10^4$ yrs.

**6)** *Accretion by predominantly small impactors.*  Small particles delivered to the disk with the inflow will grow until their mutual collision timescale is comparable to their collision timescale with the dominant large satellite in their region of the disk.  This balance yields a predicted characteristic impactor size of about a kilometer or less for the Galilean satellites (*Barr and Canup*, 2008).  However episodic collisions between large migrating satellites are possible.

**7)** *Possible primordial establishment of Laplace resonance.*  The Galilean satellites would have likely migrated inward somewhat during their formation.  Ganymede's larger mass means that it would have migrated inward at a faster rate than Europa and Io.  *Peale and Lee* (2002) have shown that it is possible that in such a situation to establish the Laplace resonance from outside-in during the satellite formation era.

**8)** *Heterogeneous accretion of Europa.*  Disk temperatures near Europa's orbit are predicted to have been too warm for ices during much of the satellite's growth, so that most of Europa would have accreted as a purely rocky object.  But during the final stages of Europa's accretion, the ice line likely moved within the satellite's orbit, so that the final ~ 10% of its accreted mass would have been a mixture of rock and ice.

TABLE 1. Satellite Properties

| | $a/R_P$ | $M_S\,(10^{25}\ \text{g})$ | $R_S\,(\text{km})$ | $\rho_S\,(\text{g cm}^{-3})$ | $C/(M_S R_S^2)$ |
|---|---|---|---|---|---|
| Io | 5.9 | 8.93 | 1822 | 3.53 | 0.378 |
| Europa | 9.4 | 4.80 | 1565 | 2.99 | 0.346 |
| Ganymede | 15.0 | 14.8 | 2631 | 1.94 | 0.312 |
| Callisto | 26.4 | 10.8 | 2410 | 1.83 | 0.355 |

Source: *Schubert et al.* (2004)

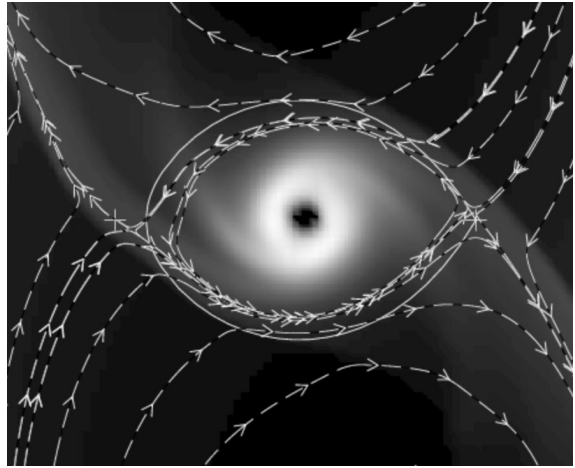

Figure 1. Example results of a three-dimensional hydrodynamical simulation of gas inflow to a Jupiter-mass planet orbiting a solar mass star at 5.2 AU, assuming $\alpha = 4 \times 10^{-3}$. Shown is a close-up on the planet's Hill sphere (solid curve), with the central star located to the right, and the planet orbiting the star in the counterclockwise direction. Brightness scales with gas density, while dashed lines and arrows indicate flow streamlines in the disk mid-plane. Material in the numerical grid zones surrounding the "planet" is assumed to be accreted, and is removed from the simulation (dark area at the center of the Hill sphere). In the mid-plane, gas flows into the Hill sphere primarily near the $L_1$ and $L_2$ Lagrange points (crosses), undergoes shocks, and settles into a prograde circumplanetary disk. From *Bate et al.*, (2003).

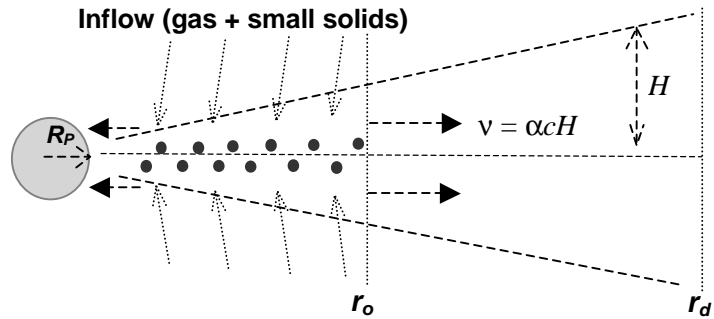

Figure 2. An inflow-produced accretion disk. Inflowing gas and solids initially achieve centrifugal force balance across a region extending from the surface of the planet out to distance $r_o$, and solids accrete into satellites throughout this region. The gas spreads viscously onto the planet and outward to a removal distance, $r_d$, with $r_d \gg r_o$. The half-thickness of the gas disk is $H \sim c/\Omega$, where $c$ is gas sound speed and $\Omega$ is orbital frequency, with $H/r \sim 0.1$. After *Canup and Ward*, (2002).

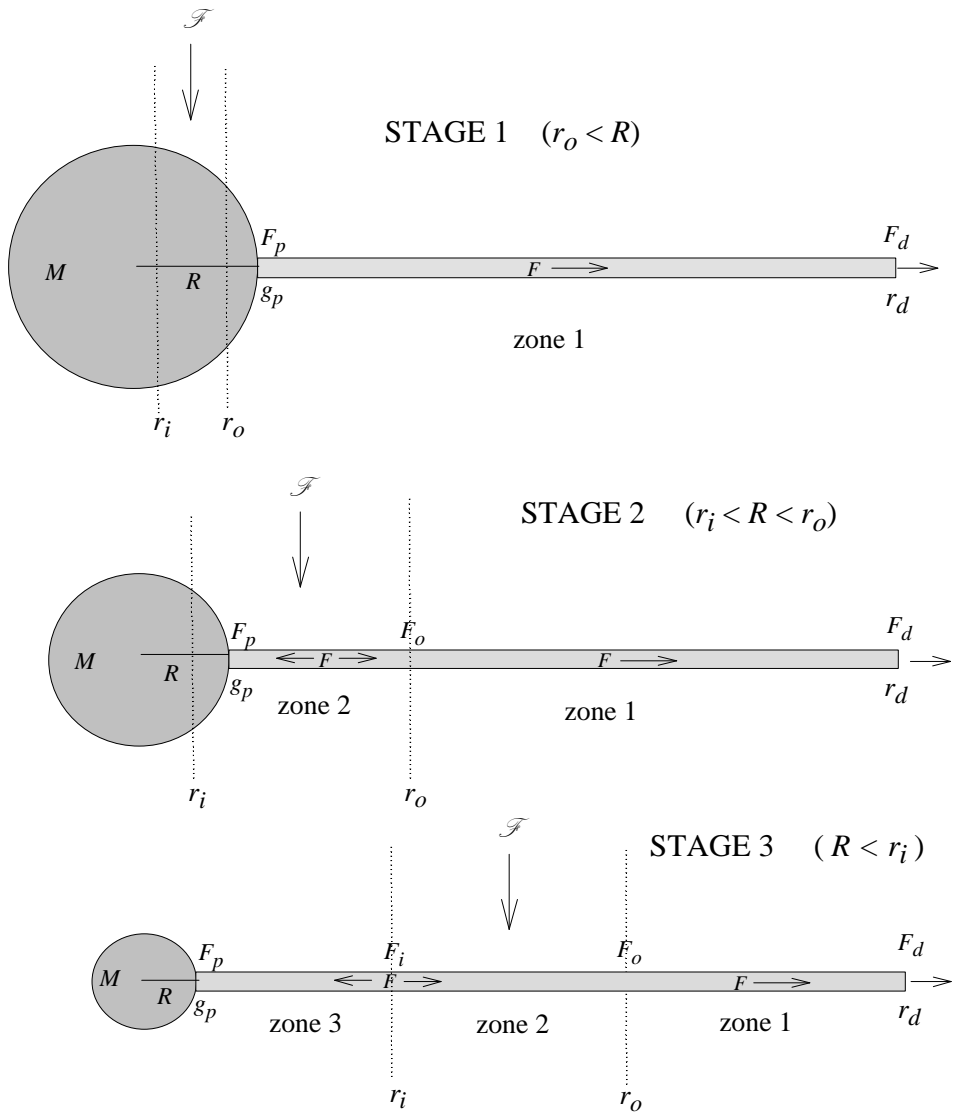

Figure 3. Possible stages of the planet-disk configuration during a Jovian planet's formation. Inflow $\mathscr{F}$ from the solar nebula occurs between $r_i$ and $r_o$. These distances grow as the planet's mass $M$ increases, while at the same time the planet's radius $R$ contracts. In stage 1, the inflow is entirely on the planet; in stage 2 it is partially on the planet and partially on the disk; in stage 3 it is entirely on the disk. The disk in-plane flux is $F$. The flux leaving the disk at its outer radius, $r_d$, is $F_d$, and is always positive; the flux at the planet, $F_p$, is the mass exchange rate between planet and disk and can be either positive (a spin-out disk) or negative (an accretion disk). In zone 2, the in-plane flux changes with radius due to the infall, while in zones 1 and 3, there is no infall and in quasi-steady state the zone fluxes remain constant.

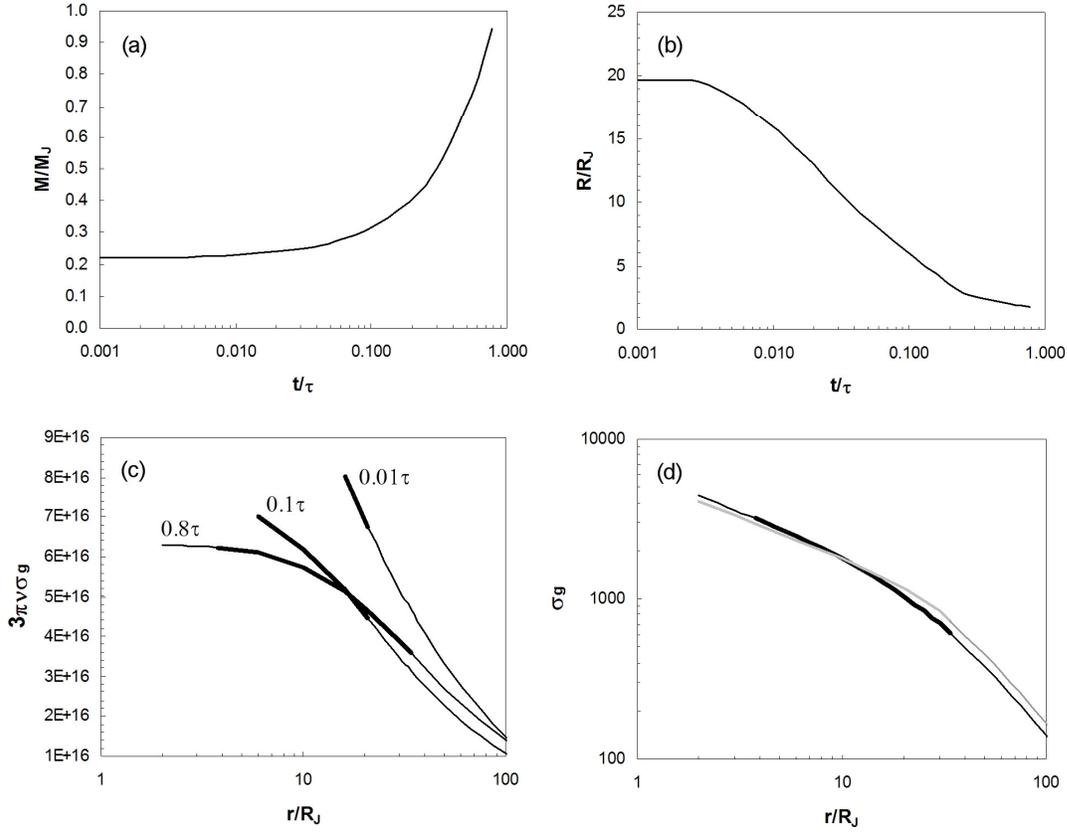

Figure 4. Circumplanetary disk properties produced during an example Jupiter growth and contraction history. (a-b) Planet mass in Jupiter masses and planet radius in Jovian radii shown as a function of time in units of the planet accretion timescale, $\tau$. This accretion and planet contraction history is from *Papaloizou and Nelson* (2005), and assumes a $5M_\oplus$ core, $\tau = 9 \times 10^5$ yr, and the opacity model of *Bell and Lin* (1994). (c) Black lines show $3\pi\sigma_g\nu$ (in cgs units) for the resulting disk produced by both inflow and spin-out, per equation (3.18), as a function of radial distance from the planet's center in Jupiter radii. The three curves correspond to three times in the planet history shown in (a) and (b) : $t = 0.01\tau$, $0.1\tau$ and $0.8\tau$. Bold sections indicate the inflow-supplied region (*i.e.* where $r_i \leq r \leq r_o$), assuming $\lambda = 0.17$, $l = 1/4$, and $r_d = 0.5R_H(M)$. (d) Sample predicted disk gas surface densities as a function of radial distance in planet radii for the $t = 0.8\tau$ curve in (c), generated using both the *Ward and Canup* model detailed in section 3.5 (black) and the *Canup and Ward* (2002) model (grey). The latter is a simpler model that considers a uniform inflow per disk area and that the disk is a pure accretion disk. Both curves assume $\alpha = 10^{-3}$, a disk temperature of 1000 K at $r = 5R_J$, and that the disk temperature decreases with $1/r$, so that $\nu \propto r^{1/2}$. Bold sections indicated the predicted inflow delivery region.

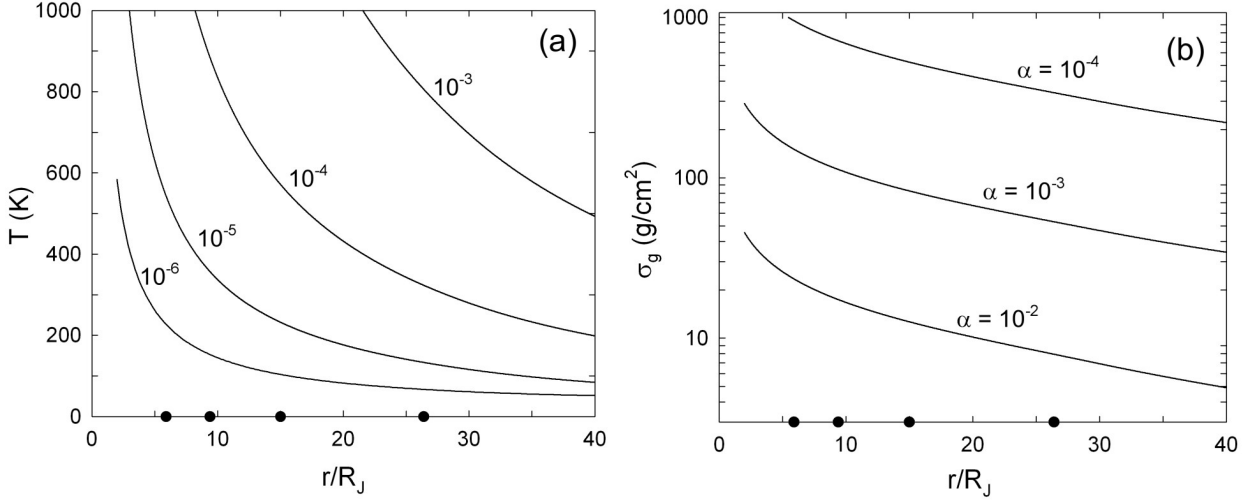

Figure 5. Properties in a protosatellite accretion disk. Current positions of the Galilean satellites are shown as circles along the *x*-axes. (a) Radial mid-plane temperature profile shown as a function of inflow rate in Earth masses per year. All four curves assume $\alpha = 0.005$, $r_0 = 30 R_P$, $r_d = 300 R_P$, $T_{neb} = 20$ K, $T_J = 500$ K, and $\kappa_R \sim \kappa_P = 1$ cm$^2$/g. As the inflow to the planet progressively slows, disk temperatures fall, with inflow rates $< 10^{-5}$ $M_\oplus$ yr$^{-1}$ yielding water ice stability in the Callisto-Ganymede region. Varying the viscosity and/or opacity for a given inflow rate affects the mid-plane temperature as $T \propto (\kappa_R / \alpha)^{1/5}$. Conditions during Galilean satellite accretion were likely intermediate to the $10^{-5}$ $M_\oplus$ yr$^{-1}$ and the $10^{-6}$ $M_\oplus$ yr$^{-1}$ curves. (b) Radial gas surface density profile for $\mathscr{F} = 10^{-5} M_\oplus$ yr$^{-1}$, shown as a function of the assumed disk viscosity $\alpha$-parameter.

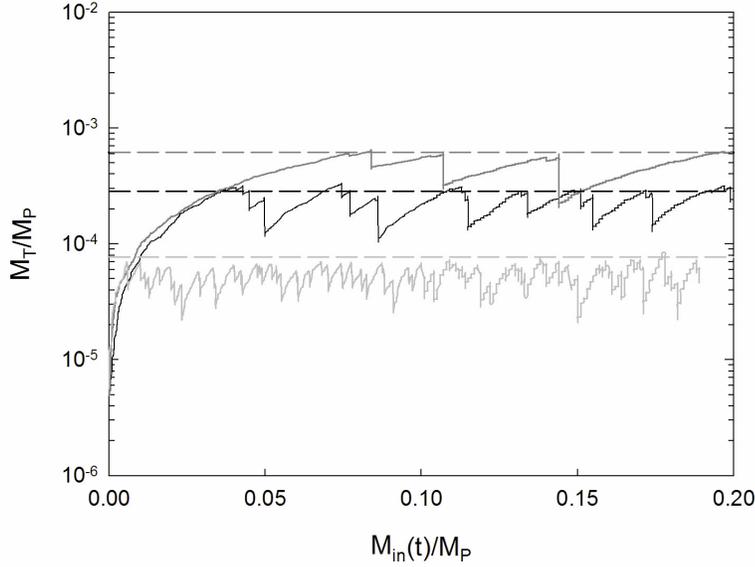

Figure 6. Results of three satellite accretion simulations that consider a time-constant rate of gas inflow to the circumjovian disk, with $\mathscr{F} = 6 \times 10^{-5}\ M_\oplus\ \text{yr}^{-1}$. The total mass in satellites ($M_T$) scaled to the planet's final mass ($M_P$) is shown vs. the cumulative total mass delivered to the disk ($M_{in}\{t\}$) in planet masses. The light grey (bottom), black (middle), and grey (top) curves correspond respectively to simulations with $(\alpha/f) = 10^{-6}$, $5 \times 10^{-5}$, and $5 \times 10^{-4}$. A satellite grows until it reaches a critical mass at which the timescale for its further growth is comparable to its Type I decay timescale. The satellite then spirals inward until it collides with Jupiter and is removed from the simulation, while in a comparable time a new, similarly sized satellite grows in its place. This process causes ($M_T/M_P$) to oscillate about a constant value. Dashed lines show analytically predicted ($M_T/M_P$) values from equation (4.15). From *Canup and Ward* (2006).

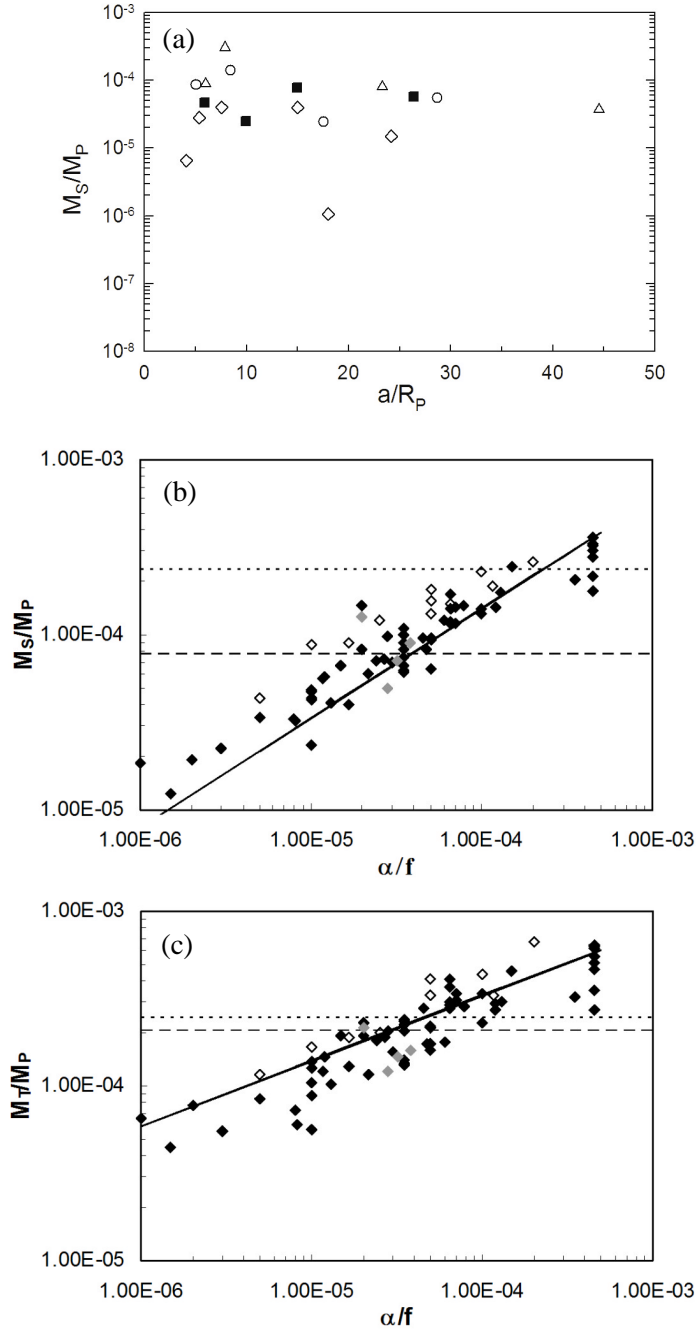

Figure 7. Results of satellite accretion simulations that considered time-dependent rates of gas and solid inflow. (a) Three final simulated satellite systems, with the Galilean satellites shown for comparison (solid squares). All 3 runs had ($r_o/R_P$) = 30, with open diamonds, circles, and triangles corresponding to runs with ($\alpha/f$) = $1.7 \times 10^{-5}$, $6.5 \times 10^{-5}$, and $5 \times 10^{-4}$, respectively. (b) The largest final satellite mass in planet masses shown as a function of ($\alpha/f$) for 75 simulations. Grey, black, and open diamonds correspond to cases with ($r_o/R_P$) = 25, 30, and 44, respectively. Dashed and dotted lines show observed values for the Galilean and Saturnian systems, while the solid line is the analytical prediction from equation (4.13). (c) The final satellite system mass in planet masses vs. ($\alpha/f$), with symbols and lines having the same meanings as in (b) and the solid line derived from equation (4.15). From *Canup and Ward* (2006).